\documentclass[twocolumn,showpacs,preprintnumbers,amsmath,amssymb]{revtex4}
\usepackage{graphicx}% Include figure files
\usepackage{dcolumn}% Align table columns on decimal point
\usepackage{bm}% bold math

\begin{document}

\preprint{APS/123-QED}

\title{The treatment of zero eigenvalues of the matrix governing the
equations of motion in \\
many-body Green's function theory.} % Force line breaks with \\

\author{P. Fr\"obrich}
\altaffiliation[Also at ]{Institut f\"ur Theoretische Physik,
Freie Universit\"at Berlin\\
Arnimallee 14, D-14195 Berlin, Germany.}

\email{froebrich@hmi.de}

%Lines break automatically or can beforced with \\
\author{P.J. Kuntz}%
 \email{kuntz@hmi.de}
\affiliation{Hahn-Meitner-Institut Berlin,
Glienicker Stra{\ss}e 100, D-14109 Berlin,
Germany}
%This line break forced with \textbackslash\textbackslash%

%\author{Charlie Author}
% \homepage{http://www.Second.institution.edu/~Charlie.Author}
%\affiliation{
%Second institution and/or address\\
%This line break forced% with \\
%}%

\date{\today}% It is always \today, today,
             %  but any date may be explicitly specified

%

%\documentclass[12pt]{article}

%\usepackage{graphicx,fancyhdr}

%

%\textheight 25cm

%\textwidth 15cm

%\topmargin -2.0cm

%\oddsidemargin -0.1cm

%\renewcommand{\baselinestretch}{1.3}

%\newcommand{\smfrac}[2]{\mbox{$\frac{#1}{#2}$}}

%\begin{document}

%\begin{abstract}

%\end{abstract}

%\tableofcontents

\def\K{\mathord{\cal K}}

\def\la{\langle}

\def\ra{\rangle}

\def\ltsim{\mathop{\,<\kern-1.05em\lower1.ex\hbox{$\sim$}\,}}

\def\gtsim{\mathop{\,>\kern-1.05em\lower1.ex\hbox{$\sim$}\,}}

%\pagestyle{empty}

% Draft: 11 (June 02, 2004)

\begin{abstract}
The spectral theorem of many-body Green's function theory relates
thermodynamic correlations to Green's functions. More often than not, the
matrix $\mathbf \Gamma$ governing the equations of motion
has zero eigenvalues. In this case, the standard text-book approach
requires both commutator and anti-commutator Green's functions to
obtain equations for that part of the correlation
which does not lie in the null space of the matrix. In this paper,
we show that this procedure fails if the projector onto the null
space is dependent on the momentum vector, $\bf k$. We propose an
alternative formulation of the theory in terms of the non-null
space alone and we show that a solution is possible if one can find a
{\em momentum-independent} projector onto some {\em subspace} of the
non-null space. To do this, we
enlist the aid of the
singular value decomposition (SVD) of $\mathbf \Gamma$ in order
to project out the null space, thus reducing the size of the matrix
and eliminating the need for the anti-commutator Green's function.
We extend our previous work\cite{FK03}, dealing with a
ferromagnetic Heisenberg
monolayer and a momentum-independent projector onto the null space,
to models where both
multilayer films and a {\em momentum-dependent} projector are
considered. We develop the numerical methods capable of handling these
cases and offer a computational algorithm
that should be
applicable to any similar problem arising in Green's function theory.
\end{abstract}


% PACS, the Physics and Astronomy

% Classification Scheme.

%\keywords{Suggested keywords}%Use showkeys class option if keyword

                              %display desired

\maketitle

\section{Introduction}
The spectral theorem in many-body
Green's function (GF) theory relates thermodynamic
correlations to Green's functions, thus providing equations
which, when iterated to self-consistency, allow the computation of
the expectation values of the operators from which the Green's functions
are constructed. Each Green's function
can be expressed in terms of an inhomogeneity of the equations of motion
and higher-order
Green's functions. Each higher-order GF can in turn be expressed
in terms of yet another
inhomogeneity and even higher-order Green's functions, and so on.
Truncation of this infinite, exact hierarchy seldom occurs
naturally. It is usually brought about
through a {\em decoupling approximation}, whereby
the GFs of some order are approximated as linear
combinations of lower-order functions which have already appeared in
the hierarchy.  This leads to a closed system of linear equations,
the so-called
{\em equations of motion}, which relate the Green's functions to the
inhomogeneities. Anticipating results from the detailed exposition
in the next section, we may write the equation of motion in compact
matrix notation:
\begin{equation}
   (\omega{\bf 1}-{\bf \Gamma}){\bf G} = {\bf A}. \label{emotion}
\end{equation}
Here, $\bf G$ is a vector whose components are the Green's functions,
$\bf A$ is a vector of associated inhomogeneities, and $\mathbf \Gamma$
is the matrix (in general unsymmetric) containing the coefficients
obtained via the truncation.

Each
Green's function has poles at all the eigenvalues, $\omega_i$, of the
matrix $\mathbf \Gamma$. As we shall show in detail in the overview
section below,
the spectral theorem associates a vector of
correlation functions, $\bf C$, to quantities at these poles, and
thereby provides a route
from the inhomogeneity vector to the vector of correlation functions.
In fact, this relationship
can be expressed compactly\cite{FJKE00} by a
multiplication of the vector $\bf A$ by a matrix constructed from the
eigenvalues, $\omega_i$,
and the right and left matrices of eigenvectors, $\bf R,L$ of
the matrix $\mathbf \Gamma$:
\begin{eqnarray}
   \bf C &=& {\bf R} {\cal E}\bf LA,               \label{erela} \\
   {\bf L} {\mathbf \Gamma} {\bf R} &=& {\mathbf \Omega},
\end{eqnarray}
where $\bf \Omega $ is a diagonal matrix having
the eigenvalues $\omega_i$ on the principal diagonal and $\cal E$
is a related diagonal matrix with
elements ${\cal E}_{ij}=\delta_{ij}/(e^{\beta\omega_i}-1)\ $,
$\beta$ being $1/(kT)$. Since
$\bf C$, $\bf A$, and $\mathbf \Gamma$ all depend on the set of expectation
values of the operators used in constructing the Green's functions,
Eq.~\ref{erela} can be solved by varying the expectation values until
the left and right hand sides of the equation are equal. Usually,
the equation holds in momentum space, so a Fourier transform
to coordinate space (an integration over the momentum $\bf k$)
is also required; i.e. a set of integral equations has to be
solved self-consistently.

This procedure, while somewhat complicated to derive,
is straightforward to apply, unless, as very often happens,
the matrix $\mathbf \Gamma$ has a null space, i.e. there are
eigenvalues of value zero.  In that case,
a naive use of Eq.~\ref{erela} would involve a division by
zero in the evaluation of $\cal E$. The standard
textbook procedure for handling the null space demands
a knowledge of the anti-commutator GF in addition to the
commutator GF from which Eq.~\ref{erela} was derived (see
references \cite{ST65,RG71} and textbooks \cite{Nol86,GHE01}). This
leads to additional terms in Eq.~\ref{erela} which restrict
the GF method to the evaluation not of the complete vector
$\bf C$, but only to that part of $\bf C$ which lies in the
non-null space of the matrix $\mathbf \Gamma$. While this in
itself is not necessarily a hindrance, it may become one if
the projector onto the null space is dependent upon
the momentum $\bf k$. In this
situation, even the modified equation Eq.~\ref{erela} fails
(see the following section);
i.e. the standard textbook solution is in this case no solution.

We propose here a new method that addresses this problem. It
exploits the singular value
decomposition (SVD) of the matrix governing the equations of motion in
the many-body Green's function theory
in order to treat the problems arising from the null space
of the matrix.  At each point in momentum space,
the SVD effects a transformation to a smaller number of Green's
functions having no associated null space, so that the corresponding
correlations can be obtained from the spectral theorem in the
straightforward manner of Eq.~\ref{erela}. This has the double
advantage of both eliminating the need
for the anti-commutator GF and
reducing the size of the $\mathbf \Gamma$-matrix, thereby also
reducing the number of coupled equations needed to iterate to
consistency. This idea was the subject of an earlier paper \cite{FK03}
on a ferromagnetic Heisenberg monolayer with single-ion anisotropy.
That system is also solvable by the standard procedure because the
projector onto the null
space is momentum independent. As such, it is an inadequate model
with which to demonstrate the effect of a momentum-dependent projector.

In the present paper, we treat as an example a model for Heisenberg thin films
with {\em exchange} anisotropy, which {\em does} lead to a momentum-dependent
projector.
Note that the SVD of $\mathbf \Gamma$ is itself dependent
upon $\bf k$ and so, while having the advantages mentioned above, does not
solve the basic problem automatically. Nevertheless, it does allow one to find
some SVD singular vectors ({\em vide infra}) which {\em are} in fact independent
of $\bf k$, thus providing equations which {\em can} circumvent the
problem of a momentum-dependent projector onto the null space.
In connection with our
method, there are a number of non-trivial numerical difficulties
which arise and
which become more acute when treating films with more than one layer.
In order to confront these problems here,
we extend our previous work\cite{FK03} on the monolayer to multilayer films and
we describe a
numerical procedure which surmounts these difficulties.

In the following section, we give an overview of the essential
equations of
GF theory needed to explicate our new method, we state the precise
nature of the problem, and we describe our proposal for solving it.
The section after that outlines the numerical problems which arise and
provides algorithms to solve them. Following that, we present the model
for ferromagnetic Heisenberg thin films with exchange anisotropy
as an illustrative
example: we present an algebraic exposition of the model for a 3-layer
film, which is easy
to extend to any number of layers. A detailed study of
the monolayer, for which all equations can be obtained analytically,
reveals the structure in the singular vectors from the SVD
and the eigenvectors of $\mathbf\Gamma$, which
harks back to the structure of the $\mathbf\Gamma$-matrix.
Some algebraic properties of the multilayers are then deduced from
the structures found in the analysis of the monolayer.
In the next section of the paper, we present some numerical
results as an illustration. In the penultimate section we
offer some remarks on how to
use the new method to aid in the design of an efficient numerical
algorithm. The last section contains a discussion and
summary.

\section{Overview}
\subsection{The standard formulation}
In this section, we supply all of the formulae necessary to make the
paper self-contained and to allow the reader to understand and make
use of the computational algorithm summarized in the discussion section.

We start with the definition of the retarded Green's functions, which
we shall use exclusively in this work:
\begin{eqnarray}
G^{\alpha}_{ij,\eta}(t-t')
  &=&-i\Theta(t-t')\langle[A_i(t),B_j(t')]_{\eta}\rangle^{\alpha}
\nonumber\\
&=&\langle\langle A_i(t);B_j(t')\rangle\rangle^{\alpha},
\label{1}
\end{eqnarray}
where $\Theta(t-t')$ is equal $1$ for $t>t'$ and $0$  for $t<t'$ .
$A_i(t)$ and $B_j(t')$ are operators in the Heisenberg picture,
and $i,j$ are lattice site
indices. $\eta=\pm1$ denotes anticommutator or commutator GF's,
$\langle ... \rangle=Tr(...e^{-\beta \cal H})/Tr(e^{-\beta\cal H})$ is the
thermodynamic expectation value, and $\cal H$ the Hamiltonian for the system
under investigation. The GFs have a label $\alpha$ because for
multidimensional problems, GFs are required for several different
operators $A$ and $B$. At this point, one need not be more specific.

Taking the time derivative of equation (\ref{1}) and performing a Fourier
transform to energy space, one obtains an exact equation of motion
\begin{equation}
\omega\la\la A_i;B_j\ra\ra^{\alpha}_{\eta,\omega}
  =\la [A_i,B_j]_\eta\ra^{\alpha}+\la\la
[A_i,{\cal H}]_-;B_j\ra\ra^{\alpha}_{\eta,\omega}. \label{2}
\end{equation}
Repeated application of the equation of motion to the higher-order Green's
functions appearing on the right-hand side of Eq.~\ref{2}
results in an
infinite hierarchy of equations. In order to obtain a solvable closed set of
equations, the hierarchy has to be terminated, usually by a decoupling
procedure which, when restricted to the lowest-order functions, leads
to the set of linear relations

\begin{equation}
\la\la [A_i,{\cal H}]_-;B_j\ra\ra^{\alpha}_\eta
 \simeq \sum_{\beta,m}{\bf \Gamma}^{\alpha \beta}_{im} \la\la A_m;B_j
 \ra\ra^{\beta}_\eta\ ,
\label{3}
\end{equation}
where $\mathbf \Gamma$ is a
matrix (in general {\em unsymmetric}) expressing the higher-order Green's
functions in terms of linear combinations of lower-order ones.

The lattice site indices can be eliminated by a Fourier transform
to momentum space and the labels $\alpha$ can be suppressed by
writing the
equation of motion in compact matrix notation, where the vectors have
components indexed by the labels $\alpha$:
\begin{equation}
(\omega {\bf 1}-{\bf \Gamma}) {\bf G}_\eta = {\bf A}_\eta,
\label{4}
\end{equation}
where ${\bf 1}$ is the unit matrix, and the inhomogeneity vector has
components ${A}^\alpha _\eta =\la [A,B]_\eta\ra^{\alpha}$.
Note that ${\bf G}_\eta$ depends on energy
and momentum (${\bf G}_\eta={\bf G}_\eta(\omega,\bf k)$) and that
${\bf A}_{+1}$ depends upon the momentum $\bf k$, whereas ${\bf A}_{-1}$
does not.

It is now convenient to
introduce the notation of the eigenvector method of
reference \cite{FJKE00}, since it is
particularly suitable for the multi-dimensional problems in which
many zero eigenvalues are likely to appear.
One starts by diagonalizing the  matrix
${\bf \Gamma}$
\begin{equation}
{\bf L\Gamma R}={\bf \Omega},
\label{5}
\end{equation}
where $\bf \Omega$ is the diagonal matrix of $N$ eigenvalues,
$\omega_\tau\  ({\tau=1,..., N})$, $N_0$ of which are zero and
$(N-N_0)$ are non-zero.
 The matrix ${\bf R}$ contains the right eigenvectors as columns and its
inverse
${\bf L}={\bf R}^{-1}$ contains
the left eigenvectors as rows. ${\bf L}$ is
constructed such that ${\bf LR}={\bf RL}={\bf 1}$.
Multiplying equation (\ref{4}) from the left by
${\bf L}$, inserting ${\bf 1}={\bf RL}$, and defining new vectors
${\cal G}_\eta={\bf LG}_\eta$ and ${\cal A}_\eta={\bf LA}_\eta$
one finds
\begin{equation}
(\omega{\bf 1}-{\bf \Omega}){\cal G}_\eta={\cal A}_\eta.
\label{6}
\end{equation}
The crucial point is that each of the components $\tau$ of this Green's
function vector has but a single pole
\begin{equation}
{(\cal G}_\eta)_\tau=\frac{({\cal A}_\eta)_\tau}{\omega-\omega_\tau}\ .
\label{7}
\end{equation}
This allows the standard spectral theorem, see e.g.
\cite{Nol86,GHE01}, to be applied
to each component of the Green's function vector {\em separately}:
One can then define the correlation vector
${\cal C}={\bf LC_{\bf k}}$, where ${\bf C_{\bf k}}=\la BA\ra$
is the vector of correlations
associated with ${\bf G}_\eta$ (the index $\bf k$ is added
to emphasize that one is in momentum space).

For the commutator functions ($\eta=-1$),
the $N-N_0$ components $\tau$ for $\omega_\tau \neq 0$,
$({\cal C}^1)_\tau$ (the upper index
1 refers to the non-null space) , are
\begin{eqnarray}
({\cal C}^1)_\tau&=&\frac{i}{2\pi}\lim_{\delta\rightarrow
0}\int_{-\infty}^{\infty} d\omega \frac{({\cal G}_{-1}(\omega+i\delta)-{\cal
G}_{-1}(\omega-i\delta))_\tau}{e^{\beta\omega_\tau}-1}\nonumber\\&=&\frac{({
\cal A}_{-1})_\tau}{e^{\beta\omega_\tau}-1}\ .
\label{8}
\end{eqnarray}
This equation cannot be used to define
the $N_0$ components of ${\cal C}^0$ (the upper index 0 refers
to the null space)
corresponding to $\omega_\tau=0$ because of the zero in the denominator.
Instead, one must enlist the help of the {\em anti-commutator}
Green's function\cite{ST65,RG71,Nol86,GHE01}:
\begin{equation}
({\cal C}^0)_{\tau_0}=\lim_{\omega\rightarrow 0}\frac{1}{2}\omega
({\cal G}_{\eta=+1})_{\tau_0}.
\label{9}
\end{equation}
The components of ${\cal C}^0$, indexed by $\tau_0$, can be simplified by using the relation between the
commutator and anti-commutator inhomogeneities,
${\bf A}_{+1}={\bf A}_{-1}+2{\bf C_{\bf k}}$. This yields, for
$\omega_{\tau_0}=0$,
\begin{eqnarray}
({\cal C}^0)_{\tau_0}&=&\frac{1}{2}\lim_{\omega\rightarrow 0}
\frac{\omega({\cal A}_{+1})_{\tau_0}}{\omega-\omega_{\tau_0}} =
\frac{1}{2}({\cal A}_{+1})_{\tau_0}\nonumber\\
&=&\frac{1}{2}({\bf L}^0({\bf A}_{-1}+2{\bf C_{\bf k}}))_{\tau_0}
=({\bf L}^0{\bf C_{\bf k}})_{\tau_0},
\label{10}
\end{eqnarray}
where we have used the regularity condition\cite{FJKE00},
${\bf L}^0{\bf A}_{-1}=0$,
which derives from the fact that the
commutator Green's function is regular at the origin:
\begin{equation}
\lim_{\omega\rightarrow 0}{\bf L}^0(\omega{\bf 1}-{\bf \Gamma}){\bf
G}_{-1}={\bf L}^0{\bf A_{-1}}=0.
\end{equation}

Again, we use superscripts 0 and 1 to
denote the vectors belonging to zero and non-zero eigenvalues,
respectively. The right and left eigenvectors and the correlation vectors may
then be partitioned as
\begin{equation}
{\bf R} = ({\bf R}^1 \ {\bf R}^0)\ \ \ \ \
   {\bf L} = \left( \begin{array}{c}
                         {\bf L}^1  \\ {\bf L}^0
                                           \end{array}      \right)\ \ \ \ \ \
                   {\cal C} = \left( \begin{array}{c}
                         {\cal C}^1  \\ {\cal C}^0
                    \end{array}      \right)\ ,
\label{11}
\end{equation}
where the correlation vectors from equations~(\ref{8}) and~(\ref{10}) are
then ${\cal C}^0 = {\bf L}^0{\bf C_{\bf k}}$
and ${\cal C}^1 = {\cal E}^1 {\bf L}^1 {\bf A_{-1}}$,
and ${\cal E}^1$ is the $(N-N_0)\times(N-N_0)$ matrix with
$1/(e^{\beta\omega_\tau}-1)\ $
on the diagonal.

Multiplying the correlation vector $\cal C$ from the left by $\bf R$ yields a
compact matrix equation for the original correlation
vector~$\bf C_{\bf k}$:
\begin{equation}
(1-{\bf R}^0{\bf L}^0){\bf C_{k}}={\bf R}^1{\cal E}^1{\bf L}^1{\bf A}.
\label{erelarolo}
\end{equation}
In the above equation and for the rest of this paper,
an inhomogeneity without a subscript
$\eta$ always refers to the case $\eta=-1$, i.e. ${\bf A}\equiv{\bf A}_{-1}$.
Eq.~\ref{erelarolo} is in momentum space
and the coupled system of integral equations obtained by  Fourier
transformation to coordinate space has to be solved self-consistently.
Usually one is interested only in the diagonal correlations; i.e. one has to
perform an integration over ${\bf k}$ in the first Brillouin zone
\begin{equation}
{\bf C}=\int d{\bf k} {\bf C_{k}},
\label{13}
\end{equation}
where the ${\bf C}$ without the subscript denotes the vector of diagonal
correlations in coordinate space.

\subsection{The problem and a proposal for solving it.}
The quantity $(1-{\bf R}^0{\bf L}^0)$ functions as a
projection operator onto the non-null
space, and as such has no inverse, so that ${\bf C_{\bf k}}$ cannot be extracted
from Eq.~\ref{erelarolo}. This tells us that the Green's function
method as formulated here can retrieve only {\em part}
of the full correlation vector in coordinate space.

In cases where the projector is momentum-independent,
which is the case in most of our previous
work \cite{FJKE00,FK03,EFJK99,FJK00,FKS02,HFKTJ02},
it is not necessary to know
the complete ${\bf C_{\bf k}}$, because one can take the projector
outside the ${\bf k}$-integration, solving
the resulting equation self-consistently in coordinate space:
\begin{equation}
(1-{\bf R}^0{\bf L}^0){\bf C}=\int d{\bf k}{\bf R}^1{\cal E}^1{\bf L}^1{\bf A}.
\label{14}
\end{equation}
The matrix elements of ${\mathbf \Gamma}$, the inhomogeneity vector $\bf A$
and the correlation vector
${\bf C}$ in coordinate space depend only on the
thermodynamic expectation values
(magnetizations and their moments), which are the variables
in terms of which the system
of equations (\ref{14}) is solved. This procedure was
followed in most of our previous
work \cite{EFJK99,FJK00,FKS02,HFKTJ02,FJKE00}.

Should, however the null-space projector depend on {\bf k}, the above
formulation cannot be used to solve for the expectation values because there is
no expression for ${\bf C_{\bf k}}$ available; i.e. the standard procedure fails.
A way around this is to transform the Green's functions so as to eliminate the
components lying in the null space. The spectral theorem then leads to a
working equation in terms of a correlation vector which lies in the non-null
space.

The tool for finding the necessary transformation is the singular value
decomposition (SVD) \cite{PFTV89,Golub} of the ${\bf \Gamma}$-matrix:
\begin{equation}
{\bf \Gamma}={\bf UW}\tilde{{\bf V}}.
\label{esvd}
\end{equation}
The matrix  ${\bf W}$ is a diagonal matrix whose elements are the
singular values, which are $\geq 0$ and ${\bf U}$ and ${\bf V}$ are
orthonormal matrices containing the singular vectors:
$\tilde{\bf U}{\bf U}={\bf 1}$ and
$\tilde{\bf V}{\bf V}={\bf 1}$, where $\tilde{\bf V}$ denotes the
transpose of ${\bf V}$.
Note that the ${\bf \Gamma}$ matrix is fully determined by the
non-zero singular values and their corresponding singular vectors alone:
\begin{equation}
{\bf \Gamma}={\bf UW}\tilde{\bf V}={\bf uw}\tilde{\bf v}.
\label{esvdnz}
\end{equation}
$\bf u$ and ${\bf v}$ are $N\times (N-N_0)$ matrices
obtained from $\bf U$ and $\bf V$ by omitting the columns corresponding
to the zero singular values.
The matrix ${\bf w}$ is the $(N-N_0)\times(N-N_0)$ diagonal matrix
with the non-zero singular values on the diagonal.
The remaining $N\times N_0$ matrices associated with the null space are
denoted by ${\bf u_0}$ and ${\bf v_0}$. The $N\times N$ matrices
${\bf v\tilde{v}}$ and ${\bf v}_0\tilde{\bf v}_0$ are
projectors onto the non-null
space and null space of ${\bf \Gamma}$, respectively.
The sum of these two projectors spans the complete space:
${\bf v}\tilde{\bf v}+{\bf v}_0\tilde{\bf v}_0=\bf 1$.
It should be borne in
mind that although ${\bf R}^0{\bf L}^0$ also behaves as a projector onto the
null space, one cannot identify ${\bf v}_0$ with ${\bf R}^0$
or $\tilde{{\bf v}}_0$ with ${\bf L}^0$, since  ${\bf R}^0$ and ${\bf L}^0$
result from diagonalization of the unsymmetric matrix ${\bf \Gamma}$.
In fact we see from Eq.~\ref{esvdnz} that ${\bf L}^0$ must lie
in the space spanned by ${\bf u}_0$ so that
$({\bf L}^0{\bf u}){\bf w}\tilde{\bf v}={\bf L}^0 \mathbf \Gamma=0$ and
similarly ${\bf R}^0$ must lie in the space spanned by ${\bf v}_0$.
Note that ${\bf V}$ and ${\bf U}$ are matrices of
the eigenvectors of the
symmetric matrices $\tilde{\bf \Gamma}{\bf \Gamma}$ and
${\bf \Gamma}\tilde{\bf \Gamma}$, respectively; the eigenvalues of
both of these matrices are the squares $w_i^2$ of the singular values of
$\mathbf\Gamma$.

The crucial point is that the dimension of the equations of motion can
be reduced by the number of zero singular values, which is equal to the
number of zero eigenvalues of $\mathbf\Gamma$, by applying the following
transformations
\begin{eqnarray}
{\bf \gamma}&=&\tilde{\bf v}{\bf \Gamma}{\bf v},\label{gamtrans} \\
{\bf g}&=&\tilde{\bf v}{\bf G}, \label{gtrans} \\
{\bf a}&=&\tilde{\bf v}{\bf A},\label{atrans} \\
{\bf c}&=&\tilde{\bf v}{\bf C_{\bf k}}. \label{ctrans}
\end{eqnarray}
This can be shown by multiplying Eq.~\ref{emotion} from the left by
$\tilde{\bf v}$ and using the identity (recall
that $\tilde{\bf v}{\bf v}=1$)
${\mathbf \Gamma}{\bf v}\tilde{\bf v}
            ={\bf uw}\tilde{\bf v}{\bf v}\tilde{\bf v}={\mathbf\Gamma}$:
\begin{eqnarray}
   \tilde{\bf v}(\omega{\bf 1}-{\mathbf \Gamma}
   {\bf v}\tilde{\bf v}){\bf G}
     &=& \tilde{\bf v}{\bf A} \nonumber \\
    (\omega {\bf 1} - \tilde{\bf v}{\mathbf \Gamma}{\bf v})\tilde{\bf v}{\bf G}
     &=& \tilde{\bf v}{\bf A} \nonumber \\
(\omega{\bf 1}-{\bf \gamma}){\bf g}&=&{\bf a}.
\label{emotionsvd}
\end{eqnarray}
Here again, the eigenvector method can be used. Matrices
${\bf l}={\bf L}^1{\bf v}$ and ${\bf r}={\bf \tilde{v}R}^1$
diagonalize the ${\mathbf\gamma}$-matrix,
${\bf l}{\mathbf\gamma}{\bf r}={\mathbf\omega}^1$, where
${\mathbf\omega}^1$ is identical to the matrix of non-zero eigenvalues
of the full $\mathbf \Gamma$-matrix (see Appendix A).
The spectral theorem applied to the equation of motion with the
matrix ${\bf \gamma}$, which now has no zero eigenvalues,
yields the equation
for the correlations ${\bf c}=\tilde{\bf v}{\bf C_{k}}$
in momentum space:
\begin{equation}
{\bf c}={\bf r}{\cal E}^1{\bf la },
\label{esvdcor}
\end{equation}
where the $(N-N_0)\times(N-N_0)$ diagonal matrix ${\cal E}^1$ has the same
elements as before,
$1/(e^{\beta\omega_\tau}-1)$.

In order to determine the correlations in coordinate space,
one has to perform a Fourier transform and then self-consistently solve the
system of integral equations
\begin{equation}
0=\int d{\bf k}({\bf r}{\cal E}^1{\bf l}\tilde{\bf v}{\bf A}-\tilde{\bf v}{\bf
C_{\bf k}})\ ,
\label{eworking}
\end{equation}
where ${\bf A}$ and ${\bf C_{\bf k}}$ are the inhomogeneity and
correlation vectors
of the original problem, and the integration is over the first Brillouin zone.
As pointed out in the introduction, Eq.~\ref{eworking} still has the problem
that the row-vectors in the matrix $\tilde{\bf v}$ are in general dependent
on the momentum; however, for the model we have investigated, it is
possible to find some row-vectors which are indeed independent of $\bf k$.
Only these rows in Eq.~\ref{eworking} may be used as working equations,
for only then can
those row-vectors in the last term on the rhs  of Eq.~\ref{eworking}
be taken outside the integral, allowing the term to be evaluated from
the correlation vector in coordinate space:
\begin{equation}
  \int d{\bf k}\ \tilde{\bf v}_j{\bf C}_{\bf k} =
  \tilde{\bf v}_j \int d{\bf k}\ {\bf C}_{\bf k} =
  \tilde{\bf v}_j {\bf C},
  \label{ecterm}
\end{equation}
where the index $j$ labels one of the $\bf k$-independent row-vectors.

The above procedure formally solves the
problem of the null space, but there remain two non-trivial problems born of
the need to use {\em numerical} methods to obtain
the vectors $\tilde{\bf v}$ and $\tilde{\bf v}_0$
in most cases.
A precise statement of these problems and a description
of the procedures needed to solve them are
the subject of the next section.

\section{Methods for solution}
Here we describe in detail the numerical difficulties which arise
because of the arbitrariness associated with any numerical determination
of the singular vectors: vectors belonging to degenerate singular
values are determined only up to an orthogonal transformation and the
phases of non-degenerate vectors are not fixed\cite{Golub}.
This arbitrariness may be removed by a smoothing
procedure applied either to the SVD vectors as they are,
(i.e. {\em untreated} vectors), or to vectors which
have been previously subjected (optionally) to a labelling procedure.
The somewhat lengthy description here will be shortened to a
recipe given later in the discussion section.
\subsection{The numerical difficulties}
When the row vectors in $\tilde{\bf v}$ are
determined {\em numerically}, they are unique
only up to a sign change or,
in the case of degenerate singular values, an orthogonal transformation
of the degenerate vectors: each time these vectors are computed
anew, they are in effect rotated by a random amount with respect to
vectors from the previous calculation; hence, even if the elements of
the $\mathbf \Gamma$-matrix are changed continuously (e.g. by varying the
momentum $\bf k$ upon which they depend),
the {\em computed} vectors $\tilde{\bf v}$ from the
singular value decomposition will not necessarily be continuous.
This is also true of the vectors in matrices
$\bf r$ and $\bf l$, which are obtained by a numerical diagonalization,
but these occur in Eq.~\ref{esvdcor} as factors separated only by a diagonal
matrix having the same degeneracy structure as the eigenvectors, so that the
arbitrariness in $\bf r$ and $\bf l$ cancel each other.
The vectors $\tilde{\bf v}$, however, appear alone, so
that vectors at neighbouring values of $\bf k$ will in general have
arbitrary phases. If they differ by a change of sign, the integrands
in Eq.~\ref{eworking} exhibit
discontinuities, preventing numerical evaluation
of the integrals over $\bf k$; we denote this problem by the term
{\em phase difficulty}.  If the vectors differ by an orthogonal
transformation, the individual members of Eq.~\ref{eworking} will refer
to different things at each value of $\bf k$, and no meaningful
system of equations can result; we shall call this the
{\em labelling difficulty}, for reasons which will become apparent.
Both of these difficulties can be overcome by {\em rotating} the
vectors $\tilde{\bf v}$ among themselves to obtain a new set which spans
the same space, is labelled, and has a smooth dependence on $\bf k$.
It should be emphasized that this rotation preserves the exact
nature of the vectors $\tilde{\bf v}$; they are not renormalized or
adjusted in any other way whatsoever.

\begin{figure}
\begin{center}
\protect
\includegraphics*[bb=0 0 451 697,
angle=0,clip=true,width=8.7cm]{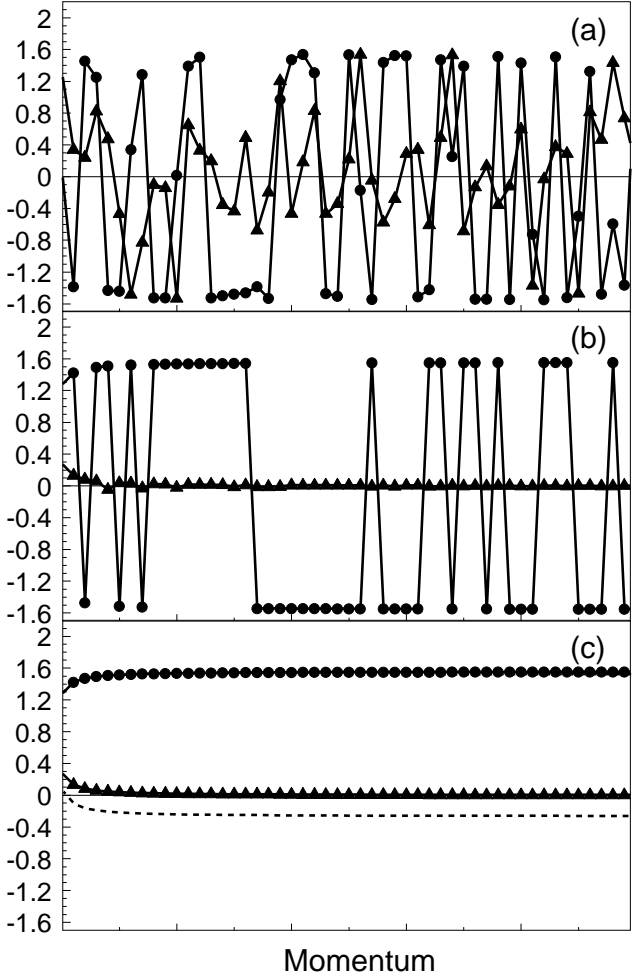}
\protect
\caption{\label{f:crazy}
         Dot products of the two $\widetilde{\bf v}_0$ vectors of a bilayer
         (see the model described in section IV)
         onto the vector (1,1,1,0,0,0) lying in the space of layer-1
         as a function of the momentum $\bf k$.
         (a) untreated vectors, (b) vectors labelled with layer index,
         (c) vectors labelled and smoothed (full line); vectors
         smoothed but not labelled (dashed line). Calculations are for a
         2-layer film with exchange energy $J=100$, exchange anisotropy
         $D=0.7$, $B^x=0.1,B^z=0$, at temperature $T=90$.
         (The reorientation temperature for the magnetization
          is $T_{\rm R}=91$).} \end{center}
\end{figure}

At this point, it is pedagogically useful
to consider a concrete example of the difficulties. To do this we
anticipate the model of section IV for a 2-layer film,
for which the matrix $\mathbf \Gamma$ has dimension 6 and the
null space has dimension 2; hence, the two null vectors from the
SVD exhibit both the labelling difficulty (because the vectors
belong to degenerate singular values) and the phase difficulty.
The first three
components of these vectors refer to layer-1 and the
second three refer to layer 2.
We calculate the dot product of each of the two vectors
$\widetilde{\bf v}_0$
with the vector $(1,1,1,0,0,0) $ (lying fully in the space of layer-1)
as a function of the momentum $\bf k$
along a line $k_x=k_y$ in the first Brillouin zone.  The numerical
computation of the dot products
suffers from both of the difficulties mentioned above, as
shown in panel $(a)$ of Fig.~\ref{f:crazy}, where the full circles
denote the dot product with the first vector and the full triangles
the dot product with the second vector. Here the words
``first'' and ``second'' refer to the order in which the vectors
are returned from the subroutine calculating the singular value
decomposition (the untreated vectors); they have no other significance.
Clearly,
the behaviour of the vectors as a function of $\bf k$ is
unacceptable, for it would prevent our successfully evaluating the integrals
numerically.
Because of the way in which we calculate the integrals in Eq.~\ref{eworking},
it is
of the utmost importance to ensure that the vectors vary smoothly with
$\bf k$ {\em before} the integral calculation is begun.
In evaluating each integral, the range of $\bf k$ is divided into
pieces and the contribution from each piece to the total integral
is summed.  Then, the pieces are chopped into smaller pieces and
the procedure repeated until the integral estimates (the sums)
from two successive subdivisions agree to within some error
tolerance.  This allows us to put the more effort into those
regions where the integrands are largest or change rapidly.
This means that two successive integrand calculations may correspond to
values of $\bf k$ far removed from each other. Thus, {\em global}
smoothness is necessary; i.e. the phases of the vectors over the whole
range of $\bf k$ must be
fixed {\em prior} to the calculation of the integrals. This is achieved
by a {\em smoothing} procedure, which we shall describe presently, but
first we address the labelling problem.

\subsection{A labelling procedure}
It may be necessary to label the vectors $\tilde{\bf v}$ and $\tilde{\bf v}_0$
either to ensure that vectors at neighbouring values of $\bf k$ refer to the
same things or to construct vectors having some specific property
(e.g. $\bf k$-independence).
It is instructive to note that, whereas the null
space vectors $\tilde{\bf v}_0$
(as delivered by the SVD) have no labels distinguishing them from each
other, the vectors $\tilde{\bf v} $ are labelled by the singular values to
which they belong, provided only that the latter are not degenerate.
Because degeneracies can arise, for some particular value of $\bf k$ say,
we cannot rely on this labelling to protect us from the kind of
discontinuities shown in panel $(a)$ of Fig.~\ref{f:crazy}.
Now the singular values are of no importance in
themselves: they serve only to separate the full space into
a null space and non-null space.  We are therefore free to take any
linear combination of the vectors $\tilde{\bf v}$ among themselves or
$\tilde{\bf v}_0$
among themselves. This freedom permits an assignment of unique labels of
our own choosing.

\begin{figure}
\begin{center}
\protect
\includegraphics*[bb=0 150 451 697,
angle=0,clip=true,width=8.7cm]{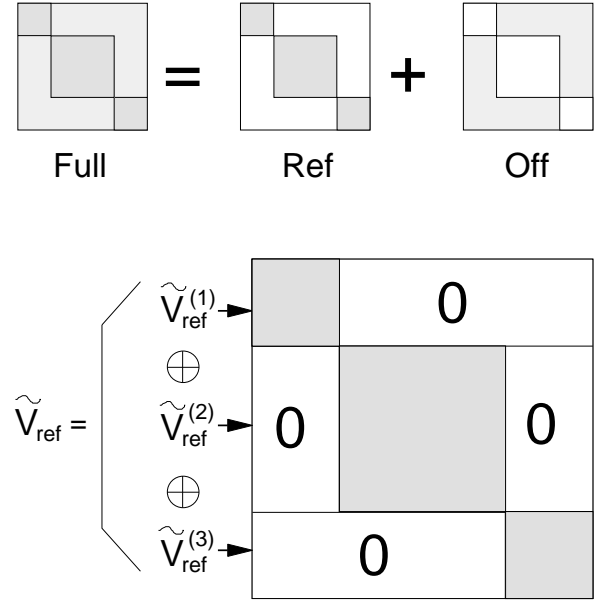}
\protect
\caption{\label{f:vrefbild}
        Full matrix $\Gamma$ expressed as a sum of the diagonal reference
        matrix ${\mathbf \Gamma}_{\rm ref}$ and the off-diagonal blocks. The
        matrix of reference vectors also has a block
        structure and is the direct
        sum of the row-vectors belonging to each block.}
\end{center}
\end{figure}

A very general approach is to generate {\em block} labels: the Green's
functions are separated into labelled sets, so that the $\mathbf \Gamma$-matrix
is partitioned into blocks characterized by the
row label index and column label index. Just how
the blocks are chosen depends upon the model under consideration. e.g. in
the model used to construct Fig.~\ref{f:crazy} ({\em vide infra}), each
block corresponds to Green's functions having the same {\em layer} index.
A set of
reference vectors, $\bf V_{\rm ref}$, can then be constructed by finding
the SVD of the associated matrix ${\mathbf \Gamma}_{\rm ref}$ which is just
the blocked $\mathbf \Gamma$-matrix with all off-diagonal blocks set to zero.
This must be done so that the reference vectors also have a block
structure: each vector has non-zero components only for the block to
which it belongs, so that it may be labelled with that block index, as
shown in Fig.~\ref{f:vrefbild}
\begin{equation}
   {\mathbf \Gamma}_{\rm ref}
   = {\bf U}_{\rm ref}{\bf W}_{\rm ref}\tilde{\bf V}_{\rm ref}.
   \label{erefvec}
\end{equation}
The set of reference vectors so constructed span the same space as
{\em all} of the original untreated vectors
($\tilde{\bf v}$ and $\tilde{\bf v}_0$ together):
${\bf V}\tilde{\bf V}={\bf V}_{\rm ref}\tilde{\bf V}_{\rm ref}$. It is
convenient to define a block-label operator in terms of the reference
vectors as
\begin{equation}
   {\bf P}_{op} :=
   \sum_{i=1}^{N_B}{\bf V}_{\rm ref}^{(i)}L(i)
   \tilde{\bf V}_{\rm ref}^{(i)},
\end{equation}
where $L(i)$ is some label for block $i$(for a specific choice, see Appendix
B), $N_B$ is the
number of blocks, and ${\bf V}_{\rm ref}^{(i)}$ is the set of reference vectors
belonging to block $i$. The matrix of ${\bf P}_{op}$ in the basis
of the singular vectors $\bf v$ (or, analogously, ${\bf v}_0$, if
these are needed) of the full $\mathbf\Gamma$ matrix
is
\begin{equation}
  {\bf P} = \tilde{\bf v}{\bf P}_{op}{\bf v} =
             \sum_{i=1}^{N_B}\tilde{\bf S}_i{\bf S}_i,
\end{equation}
where ${\bf S}_i := \sqrt{L(i)}\tilde{\bf V}_{\rm ref}^{(i)}{\bf v}$
This is equivalent to defining $\bf P$ as a product of overlap matrices
expressed in terms of a matrix of the reference vectors each multiplied
by the square root of their labels:
\begin{eqnarray}
   {\bf P} &=& \tilde{\bf S}{\bf S},  \\
   {\bf S} &=& \left[\sqrt{L(1)}\tilde{\bf V}_{\rm ref}^{(1)}\oplus \ldots
                \oplus \sqrt{L(N_B)}\tilde{\bf V}_{\rm ref}^{(N_B)}\right]
                {\bf v}.
\end{eqnarray}
See Fig.~\ref{f:vrefbild} for the meaning of the direct sum symbol $\oplus$.
We now write $\bf S$ in terms of its singular value
decomposition:
\begin{eqnarray}
   {\bf S} &=& {\bf Ly}\tilde{\bf Z},   \\
   \tilde{\bf S}{\bf S} &=& {\bf Zy}^2\tilde{\bf Z}.
\end{eqnarray}
i.e. {\bf Z} diagonalizes $\tilde{\bf S}{\bf S}$, the matrix
of the block-label operator. The singular values of {\bf S} are the
square roots of the eigenvalues of the block-label operator in the
basis $\bf v$. In other words, if each vector in $\tilde{\bf v}$ could be
brought into coincidence with one of the reference vectors, then these
vectors would have the same labels as the reference vectors.  This does
not, in general, occur, because the off-diagonal blocks of the
$\mathbf \Gamma$-matrix ensure that the reference vectors cannot all be
singular vectors of the full matrix; hence, the squares
of the singular values of {\bf S}
can be used as labels of transformed vectors, $\tilde{\bf v}_L$:
\begin{equation}
   \tilde{\bf v}_L = {\tilde{\bf Z}\tilde{\bf v}}.
\label{elabel}
\end{equation}

The matrix of the block-label operator in the basis ${\bf v}_L$
is ${\bf y}^2$, a {\em diagonal} matrix.
This tells us that we have found a rotation
(recall that $\tilde{\bf Z}{\bf Z}=\bf 1$) that associates the
vectors $\tilde{\bf v}_L$ as closely as possible with the reference vectors.
If the block-label method is to be practicable, then the values $y^2$
should lie close to the labels $L(i)$.

In the concrete example of Fig.~\ref{f:crazy}, panel $(b)$ shows the
dot products again, this time computed from vectors $\tilde{\bf v}_{0,L}$
labelled with
a layer-index.  The erratic behaviour has now disappeared: it is clear
that one of the null vectors, (the one whose dot product is nearly zero)
is now associated largely with layer~2, while the other is associated
with layer~1. Note that the discontinuities coming from the
arbitrary signs
of the vectors still remains.  This can be cured with the smoothing
procedure.

\subsection{The smoothing procedure}

At this point, it is
convenient to drop the bold-face $\bf k$, writing simply $k$ instead,
since we can, without loss of generality, discuss the methodology in
terms of 1-dimensional integrals only.  In fact, our models employ
the simple cubic lattice, where the 2-D integrals
can be reduced to 1-D integrals by exploiting the symmetry in the
terms containing $k_x$ and $k_y$. The first Brillouin zone is mapped
onto the $k$-line in the interval $0 \leq k \leq 1$. Note that $k$ is
{\em not} simply the magnitude of $\bf k$, but rather a parameter
which determines the value of all terms which depend on the momentum
$\bf k$.

The smoothing procedure is akin to the labelling method in
that untreated vectors $\tilde{\bf V}$, which we shall here
refer to as target vectors, are brought into
as close a coincidence as possible with the
reference vectors. i.e. the target vectors are {\em rotated} to match
the (fixed) reference vectors as well as possible.
Here, however, the reference vectors are sets of
standard vectors determined at various reference values of $k$
in the interval $0\leq k \leq 1$. As such, the reference vectors do
not span exactly the same space as the set of target vectors at a
neighbouring value of $k$, so a slightly different procedure must be
employed. Note that in this section, the untreated
vectors $\tilde{\bf V}$ (${\bf V}=({\bf v},{\bf v}_0)$) may
or may not be labelled vectors $\tilde{\bf V}_L$, so we shall drop the
subscript $L$ for generality. If the vectors are not labelled, the method
presented here must be powerful enough to cure the erratic behaviour shown in
panel $(a)$ of Fig.~\ref{f:crazy}, not just the sign changes shown
in panel $(b)$ of the same figure. (See the dashed line in
panel $(c)$ of the figure for such a case.)

We achieve the global smoothing in two stages. {\em Prior} to the
integral calculation, we compute sets of well-placed
reference vectors, $\{\tilde{\bf V}^{(r)}_{\rm ref}; r=1,\ldots N_r\}$
corresponding to points in the $k$-interval. These allow us to generate
by interpolation an appropriate reference vector anywhere on the
$k$-interval. Then, {\em during} the calculation of the integrands
in Eq.~\ref{eworking},
we {\em rotate} the vectors $\tilde{\bf v}$ at a
particular $k$ so as to match as closely as possible the
interpolated reference vectors appropriate for that $k$.

To construct the sets of reference vectors, we start by obtaining the
vectors $\tilde{\bf V}_{\rm ref}(k_0)$ at some point $k_0$, taking these to be
{\em primary} reference vectors. Moving away from $k_0$ a small
distance (e.g. a tenth of the $k-$range), we calculate the vectors
$\tilde{\bf V}(k_1)$ at the trial point $k_1$ and from these,
we construct a
set of reference vectors $\tilde{\bf V}_{\rm ref}(k_1)$
by a procedure designed to
find the best match of $\tilde{\bf V}_{\rm ref}(k_1)$
with $\tilde{\bf V}_{\rm ref}(k_0)$. (An exposition of the
methods to find a general rotation matching two vector
spaces can be found in Appendix B.)
The overlaps of corresponding vectors in the
two sets are used as a criterion for accepting or rejecting the
trial point.  If the point is accepted, points further away are
tested until either some maximum allowed distance is reached, or
a point fails the test.  If the first trial point is rejected,
the step-size is halved to get a new test point.  In this way, a
set of {\em secondary} reference points can be found linking the
vectors over the whole $k$-interval to the primary reference vectors.
 For the work reported here, sets of about
10 to 25 reference points sufficed to ensure a very large overlap
(0.98, where 1.0 is ``perfect'') of corresponding reference
vectors at neighbouring $k$-points.

Once the sets of reference vectors are in place at the $N_r$ reference points
$\{k_i, i=0,1,2,\ldots, N_r\}$, smoothed vectors at any $k$ may be obtained
in two steps: 1) a set of reference vectors at $k$ is obtained by
interpolating between the vectors at the reference points $k_l$ and
$k_h$ which bracket $k$: $k_l \leq k \leq k_h$. 2) The vectors $\tilde{\bf v}$
(untreated or labelled) are then rotated to be as close as possible
to those in the interpolated reference set.

The computation of the interpolated reference vectors requires three steps:
\begin{enumerate}
   \item Weights are defined for the vectors at $k_l$ and $k_h$:
   \begin{eqnarray*}
   w_l &=& \cos^2{\frac{\pi}{2}\left( \frac{k-k_l}{k_h-k_l}\right)}, \\
   w_h &=& 1 - w_l.
   \end{eqnarray*}
   \item A set of interpolated approximate reference vectors at $k$,
    $\widetilde{\overline{\bf V}}_{\rm ref}(k)$ is obtained
         as a weighted sum of the vectors at $k_l$ and $k_h$:
         \begin{equation}
           \widetilde{\overline{\bf V}}_{\rm ref}(k)
            = w_l\tilde{\bf V}_{\rm ref}(k_l)+w_h
            \tilde{\bf V}_{\rm ref}(k_h).
         \end{equation}
   \item The approximate vectors are orthonormalized by a transformation
         found by diagonalizing their overlap matrix $\cal Y$:
       \begin{eqnarray*}
          {\cal Y} &:=&
          \widetilde{\overline{\bf V}}_{\rm ref}\
          \overline{\bf V}_{\rm ref}, \\
          {\bf \lambda} &= & \tilde{\bf T}{\cal Y \bf T}, \\
          {\cal Y}^{-1/2} &=& {\bf T\lambda}^{-1/2}\tilde{\bf T}, \\
          \tilde{\bf V}_{\rm ref}
           &=& {\cal Y}^{-1/2}\ \widetilde{\overline{\bf V}}_{\rm ref}.
       \end{eqnarray*}
\end{enumerate}
The overlap matrix of the transformed vectors is clearly unity:
\begin{eqnarray*}
 \tilde{\bf V}_{\rm ref}{\bf V}_{\rm ref}  &=&
  {\bf T\lambda}^{-1/2}(\tilde{\bf T}{\cal Y}{\bf T})
  \lambda^{-1/2}\tilde{\bf T}, \\
  &=& {\bf T\lambda}^{-1/2}(\lambda)\lambda^{-1/2}\tilde{\bf T}, \\
  &=& {\bf 1}.
\end{eqnarray*}
The functional dependence of the weights chosen here ensures that
the interpolated quantities are
continuous and smooth at the reference points.
We now have reference vectors for the non-null space and the null
space (${\bf V}_{ref}=({\bf v}_{ref},{\bf v}_{0,ref})$).

The procedure for matching the untreated (or labelled) vectors $\tilde{\bf v}$
with the interpolated reference
vectors $\tilde{\bf v}_{\rm ref}$ {\em during} the integrand calculation
is to seek a transformation
$\bf Q$ that rotates the target vectors among themselves so as to
achieve the best match, just as in the labelling procedure:
\begin{equation}
   \tilde{\bf v}_S = {\bf Q}\tilde{\bf v}.
\end{equation}
$\bf Q$ is obtained via the singular value decomposition of the overlap
matrix of the reference vectors with the target vectors:
\begin{eqnarray*}
    {\cal S} &:=& \tilde{\bf v}_{\rm ref}{\bf v}, \\
            &=& {\cal L}{\bf x}\tilde{\cal Z},       \\
    {\bf Q} &=& {\cal L}\tilde{\cal Z},
\end{eqnarray*}
where $\bf x$ is the diagonal matrix of singular values of the overlap
matrix, which are all very close to unity by construction. $\bf Q$
is indeed a rotation because
\begin{equation}
  \tilde{\bf Q}{\bf Q}= {\cal Z}\tilde{\cal Z} = \bf 1,
\end{equation}
the latter equality holding because the overlap matrix has no null space.
The overlap of the transformed vectors with the reference vectors
is almost the unit matrix:
\begin{equation}
    \tilde{\bf v}_{\rm ref}{\bf v}_S = {\cal SZ}\tilde{\cal L}
      = {\cal L}{\bf x}\tilde{\cal Z}{\cal Z}\tilde{\cal L}
      ={\cal L}{\bf x}\tilde{\cal L} \approx \bf 1.
\end{equation}
i.e. the new vectors are as close as possible to the reference vectors.
This procedure fixes the phases of the new vectors, because the
transformation matrix {\bf Q} is the {\em product} of the
matrices $\cal L$ and $\tilde{\cal Z}$ which stem from a single SVD
computation, so that the arbitrariness in the numerical
determination of $\cal L$ and $\tilde{\cal Z}$ is lifted.

To summarize:
the untreated vectors can be both labelled and smoothed by the
transformation
\begin{equation}
   \tilde{\bf v}_{LS} = {\bf Q}\tilde{v}_L
               ={\cal L}\tilde{\cal Z}\tilde{\bf Z}\tilde{\bf v}.
\end{equation}

In our concrete example, the effect of the
smoothing operation is seen in panel $(c)$ of
Fig.~\ref{f:crazy}. The solid lines are the result of applying the
smoothing to the labelled vectors in panel $(b)$:
$\tilde{\bf v}_{0,SL}={\cal L}\tilde{\cal Z}\tilde{\bf Z}\tilde{\bf v}_0$,
where $\tilde{\bf v}_0$ are the untreated vectors labelled only by
the singular values $0$.
The dashed lines result
from smoothing the untreated vectors in panel $(a)$:
$\tilde{\bf v}_{0,S}={\cal L}\tilde{\cal Z}\tilde{\bf v}_0$.

In our example, we have not attempted to
calibrate the primary vectors in any way, but this could
be done if desired; however, this would still be a very crude way
of labelling the vectors. A more important function of the primary
vector is to ensure that all points in the region of a given point
are related to one another. $k$ is not the only parameter determining
the elements of the $\mathbf \Gamma$-matrix; the expectation values of
the spin operators are others.  In particular, if the derivatives
of some quantity with respect to any of these parameters is needed,
then it is a good idea to retain some standard vector with which
to calibrate the primary vectors for all points in the neighbourhood
of some given point.

\section{Heisenberg Hamiltonian with exchange anisotropy}
\subsection{Algebraic formulation}
In order to illustrate the labelling and smoothing procedures
described here, we consider the model
of reference\cite{FK02}, a Heisenberg Hamiltonian with an exchange
anisotropy for thin ferromagnetic films.
In contrast to the work in reference\cite{FK03}, the
projector onto the null space here depends on the momentum. To
show the importance of the labelling procedure, we need to extend the
model of reference \cite{FK03} to {\em multilayer}
thin films.

In what follows, a composite subscript
of the type
$k_\kappa$ refers to the site $k$ in layer $\kappa$.
We shall consider only nearest neighbour interactions
in a simple cubic lattice structure.
The Hamiltonian is characterized by an exchange
interaction with strength ($J_{k_\kappa l_\lambda}>0$)
between nearest neighbour lattice sites,
a uniaxial exchange
anisotropy in the z-direction with strength ($D_{k_\kappa l_\lambda}>0$),
and an external magnetic
field ${\bf B}=(B^x,0,B^z)$ confined to the reorientation plane of
the magnetization:
\begin{eqnarray}
{\cal H}=&-&\frac{1}{2}\sum_{<k_\kappa l_\lambda>}J_{k_\kappa l_\lambda}
         (S_{k_\kappa}^-S_{l_\lambda}^++S_{k_\kappa}^zS_{l_\lambda}^z)
\nonumber\\
&-&\frac{1}{2}\sum_{<k_\kappa l_\lambda>}D_{k_\kappa l_\lambda}
          S_{k_\kappa}^zS_{l_\lambda}^z
\nonumber\\
&-&\sum_k\Big(B^x\frac{1}{2}(S_{k_\kappa}^++S_{k_\kappa}^-)
  +B^zS_{k_\kappa}^z\Big).
\label{ehamil}
\end{eqnarray}
Here the notation $S_{k_\kappa}^{\pm}=S_{k_\kappa}^x\pm iS_{k_\kappa}^y$  is
introduced, where $k_\kappa$ and $l_\lambda$ are lattice site indices
and $<k_\kappa l_\lambda>$ indicates
summation over nearest neighbours only. To fully describe magnetization
in the $xz$-plane for arbitrary spin $S$, Green's functions of
the following type are required:
\begin{equation}
   G_{r_\rho,s_\sigma}^{\alpha,mn} = \ll S_{r_\rho}^\alpha ;
   (S_{s_\sigma}^z)^m(S_{s_\sigma}^-)^n\gg, \label{egreenf}
\end{equation}
where $\alpha$ is one of $\{+,-,z\}$ and $m$ and $n$ are integers
which are determined by the spin $S$.
The equation of motion for this Green's function is
\begin{eqnarray}
   \omega G_{r_\rho,s_\sigma}^{\alpha,mn} &=& A_{r_\rho,s_\sigma}^{\alpha,mn}
   + \ll \left[S_{r_\rho}^\alpha,{\cal H}\right]_- ;
   (S_{s_\sigma}^z)^m(S_{s_\sigma}^-)^n\gg,  \nonumber \\
A_{r_\rho,s_\sigma}^{\alpha,mn}  &=& \left\la \left[ S_{r_\rho}^\alpha ,
   (S_{s_\sigma}^z)^m(S_{s_\sigma}^-)^n\right]_-\right\ra.
\end{eqnarray}
The generalized Tyablikov decoupling approximation\cite{Tya59} is now applied
to each Green's function on the rhs of the equation of motion, e.g.
\begin{equation}
   \ll S_{r_\rho}^\alpha S_{t_\tau}^\beta ;\ldots \gg
   \approx \langle S_{r_\rho}^\alpha\rangle\  G_{t_\tau,s_\sigma}^{\beta,mn}
   +\langle S_{t_\tau}^\beta\rangle\  G_{r_\rho,s_\sigma}^{\alpha,mn}.
\end{equation}
Elimination of the site indices by means of a Fourier transformation to
momentum space results in the matrix form of the equations of motion
shown in Eq.~\ref{emotion}. The components in the Green's function
vector have the superscripts $\alpha$ and $mn$ and two subscript
layer-indices and depend on energy and momentum:
$G_{\rho\sigma}^{\alpha,mn}(\omega,\bf k)$.

\subsubsection{The 3-layer model}
We now specialize the exposition to the 3-layer film, since it is
the smallest non-trivial example from which the extension to larger
films is obvious. For this case,
Eq.~\ref{emotion} has the following structure:
\begin{eqnarray}
 \left[ \omega {\bf 1} -
 \left( \begin{array}{ccccccccc}
 \Gamma_{11} & \Gamma_{12} & 0 & 0 & 0 & 0 & 0 & 0 & 0 \\
 \Gamma_{21} & \Gamma_{22} & \Gamma_{23} & 0 & 0 & 0 & 0 & 0 & 0 \\
     0       & \Gamma_{32} & \Gamma_{33} & 0 & 0 & 0 & 0 & 0 & 0 \\
%\hline
  0 & 0 & 0 & \Gamma_{11} & \Gamma_{12} &           0 & 0 & 0 & 0 \\
  0 & 0 & 0 & \Gamma_{21} & \Gamma_{22} & \Gamma_{23} & 0 & 0 & 0  \\
  0 & 0 & 0 &     0       & \Gamma_{32} & \Gamma_{33} & 0 & 0 & 0  \\
%\hline
 0 & 0 & 0 &  0 & 0 & 0 & \Gamma_{11} & \Gamma_{12} &           0  \\
 0 & 0 & 0 &  0 & 0 & 0 & \Gamma_{21} & \Gamma_{22} & \Gamma_{23}  \\
 0 & 0 & 0 &  0 & 0 & 0 &     0       & \Gamma_{32} & \Gamma_{33}
\end{array} \right) \right]
 & &   \nonumber \\
 {\Large \times}
\left( \begin{array}{c}
  G_{11} \\
  G_{21} \\
  G_{31} \\
  G_{12} \\
  G_{22} \\
  G_{32} \\
  G_{13} \\
  G_{23} \\
  G_{33}
\end{array} \right)
   =
\left( \begin{array}{c}
  A_{11} \\
    0    \\
    0    \\
    0    \\
  A_{22} \\
    0    \\
    0    \\
    0    \\
  A_{33}
\end{array} \right). \rule{1.8cm}{0cm}
   & &
\ . \label{bigamma}
\end{eqnarray}
Each of the entries in the $9\times 9$ matrix ${\mathbf \Gamma}$ is in
fact a $3\times 3$ matrix corresponding to a Green's function vector with
the same $mn$ values and layer subscripts but with superscripts
$\alpha=\{+,-,z\}$ characterizing the vector components. Using $i$ as a
layer-index, the diagonal matrix ${\mathbf \Gamma_{ii}}$ is
\begin{equation}
 {{\mathbf \Gamma}_{ii}}= \left( \begin{array}
{@{\hspace*{3mm}}c@{\hspace*{5mm}}c@{\hspace*{5mm}}c@{\hspace*{3mm}}}
\;\;\;H^z_i & 0 & -H^x_i \\ 0 & -H^z_i & \;\;\;H^x_i \\
-\frac{1}{2}\tilde{H}^x_i & \;\frac{1}{2}\tilde{H}^x_i & 0
\end{array} \right)
\ . \label{gamii}
\end{equation}
where
\begin{eqnarray}
H^z_i&=&Z_i+\la S_i^z\ra J_{ii}(q-\gamma_{\bf k})\ ,
\nonumber\\
Z_i&=&B^z
+D_{ii}q\la S^z_i\ra
+(J_{i,i+1}+D_{i,i+1})\la S_{i+1}^{z}\ra
\nonumber\\
& &+(J_{i,i-1}+D_{i,i-1})\la
S_{i-1}^{z}\ra\ ,\\
\tilde{H}^x_i&=&B^x+\la S_i^x\ra J_{ii}(q-\gamma_{\bf
k})
+J_{i,i+1}\la S_{i+1}^{x}\ra+J_{i,i-1}\la
S_{i-1}^{x}\ra \ ,
\nonumber\\
H^x_i&=&\tilde{H}^x_i-\la S_i^x\ra D_{ii}\gamma_{\bf k} \ .
\nonumber
\label{diagdef}
\end{eqnarray}
For a square lattice and a lattice constant taken to be unity,  $\gamma_{\bf
k}=2(\cos k_x+\cos k_y)$, and $q=4$ is the
number of intra-layer nearest neighbours.
The off-diagonal sub-matrices ${\bf \Gamma}_{ij}$ for $j= i\pm 1$ are of the
form
\begin{equation}
 {\bf \Gamma}_{ij} = \left( \begin{array}{ccc}
%{@{\hspace*{3mm}}c@{\hspace*{5mm}}c@{\hspace*{5mm}}c@{\hspace*{3mm}}}
-J_{ij}\la S_i^z\ra & 0 & \;\;\;(J_{ij}+D_{ij})\la S_i^x\ra \\
0 & \;\;J_{ij}\la S_i^z\ra & -(J_{ij}+D_{ij})\la S_i^x\ra \\
\frac{1}{2}J_{ij}\la S_i^x\ra &
-\frac{1}{2}J_{ij}\la S_i^x\ra & 0 \end{array} \right) \;.
\label{gamij}
\end{equation}
Unlike the elements of the $\mathbf \Gamma$-matrix, the components of
the vectors $\bf C$ and $\bf A$ are dependent on the
values of $m$ and $n$. For spin $S=1$, the values of $(m,n)$ required
are $(0,1)$ and $(1,1)$. The values of the diagonal correlations
$C_{ii}$ and $A_{ii}$ for each value of $\alpha$ are shown in
Table~\ref{t:canda}. The layer index $i$ has been omitted for brevity.
\begin{table}
   \vspace{0cm}
   \caption{\label{t:canda}Diagonal correlations and inhomogeneities
    for spin $S=1$.}
\begin{tabular}{|c|c|cc|}\hline\hline
        & $\alpha$ & $mn=(0,1)$ & $mn=(1,1)$ \\ \hline
        &     +    &   $\la 2S^z\ra$   & $-2-\la S^z\ra +3\la S^zS^z\ra$ \\
$\bf A$ &     -    &     $0$           &  $\la S^xS^x\ra$          \\
        &     $z$  &   $-\la S^x \ra $   &  $-\la S^zS^x \ra $       \\ \hline
        &     +    & $2-\la S^z \ra - \la S^zS^z \ra $ & $\la S^z \ra - \la S^zS^z \ra$  \\
$\bf C$ &     -    &     $\la S^xS^x \ra $    & $\la S^zS^xS^x \ra $     \\
        &     $z$  &   $\la S^x \ra +\la S^zS^x \ra $ & $\la S^zS^x \ra + \la S^zS^zS^x \ra $ \\  \hline \hline
\end{tabular}
\end{table}

The block-diagonal structure of the matrix in Eq.~\ref{bigamma} allows us
to write Eq.~\ref{bigamma} in terms of Green's functions and
inhomogeneities labelled by a single layer index $i$. Define singly-indexed
Green's function vectors (here $3\times 3=9$ components),their
corresponding correlations, and inhomogeneities for $\{i=1,2,3\}$.
\begin{displaymath}
\begin{array}{ccc}
   G_i = \left( \begin{array}{c}
                 G_{1i} \\
                 G_{2i} \\
                 G_{3i}
                \end{array} \right),  &
   C_i({\bf k}) = \left( \begin{array}{c}
                 C_{1i} \\
                 C_{2i} \\
                 C_{3i}
                \end{array} \right),  &
   A_i = \left( \begin{array}{c}
                 A_{1i}\delta_{1i} \\
                 A_{2i}\delta_{2i}  \\
                 A_{3i}\delta_{3i}
                \end{array} \right).
\end{array}
\end{displaymath}
The $(9\times 9)$ matrices are independent of an index:
\begin{equation}
   \Gamma = \left( \begin{array}{ccc}
                    \Gamma_{11} & \Gamma_{12} & 0 \\
                    \Gamma_{21} & \Gamma_{22} & \Gamma_{23} \\
                       0        & \Gamma_{32} & \Gamma_{33}
                   \end{array} \right). \label{gamthree}
\end{equation}
The big equations can now be replaced by 3 smaller equations of motion
and correlation equations $i=1,2,3$:
\begin{eqnarray}
   \left( \omega {\bf 1} - \Gamma \right)G_i & = & A_i, \\
({\bf 1}-{\bf R}^0{\bf L}^0)C_i({\bf k})
  &=& {\bf R}^1{\cal E}^1{\bf L}^1{A_i}.
\label{ebigc}
\end{eqnarray}

Applying the SVD to $\Gamma$ yields six vectors $\tilde{\bf v}$ in the
non-null space (each layer contributes one null vector). The corresponding
six correlations $\bf c$ are obtained by multiplying the last equation by
$\tilde{\bf v}$ and inserting
${\bf v}\tilde{\bf v}+{\bf v}_0\tilde{\bf v}_0=\bf 1$:
\begin{equation}
 \tilde{\bf v}({\bf 1}-{\bf R}^0{\bf L}^0)C_i({\bf k})
      = {\bf R}^1{\cal E}^1{\bf L}^1
      ({\bf v}\tilde{\bf v}+{\bf v}_0\tilde{\bf v}_0) A_i.
\end{equation}
Use of $\tilde{\bf v}{\bf R}^0={\bf 0}$,
${\bf L}^1{\bf v}_0={\bf 0}$ (see Appendix A),
${\bf r} = \tilde{\bf v}{\bf R}^1$, and ${\bf l}={\bf L}^1{\bf v}$
leads to
\begin{equation}
   \tilde{\bf v}C_i({\bf k}) = {\bf r}{\cal E}^1{\bf l}\tilde{\bf v}A_i,
   \label{smallc}
\end{equation}
which, after integration over $\bf k$ corresponds to Eq.~\ref{eworking}.

\subsubsection{Analytical algebraic results for the monolayer}
We now investigate in detail the monolayer for spin $S=1$,
for which many
results can be obtained analytically and reveal features which
are pertinent to the structures found for the multilayers .

The monolayer model leads to an equation of motion with the single
diagonal block ${\mathbf \Gamma}_{11}$ and the vectors ${\bf G}_{11}$,
${\bf C}_{11}$, and ${\bf A}_{11}$. For the remainder of this section,
we shall drop the subscripts $11$. The eigenvalues of $\mathbf \Gamma$
are $\{0,\epsilon_k,-\epsilon_k\}$, where
$\epsilon_k=\sqrt{H^zH^z+H^x\tilde{H^x}}$. The matrices of eigenvalues
and eigenvectors of $\mathbf \Gamma$ are then
\begin{eqnarray}
   {\bf L}{\mathbf \Gamma}{\bf R} &=& \Omega = \left(
           \begin{array}{ccc}
              0 & 0 & 0 \\
              0 & \epsilon_k & 0 \\
              0 & 0 & -\epsilon_k
           \end{array}
           \right),             \nonumber             \\[2mm]
   {\bf R} & = & \left(
         \begin{array}{ccc}
           \frac{H^x}{H^z} & \frac{-(\epsilon_k+H^z)}{\tilde{H}^x} &
           \frac{(\epsilon_k-H^z)}{\tilde{H}^x} \\
           \frac{H^x}{H^z} & \frac{(\epsilon_k-H^z)}{\tilde{H}^x} &
           \frac{-(\epsilon_k+H^z)}{\tilde{H}^x} \\
           1 & 1 & 1
         \end{array}
         \right),  \label{lreqn}                   \\[2mm]
    {\bf L} & = & \frac{1}{4\epsilon_k^2}\left(
         \begin{array}{ccc}
            2\tilde{H}^xH^z & 2\tilde{H}^xH^z &  4H^zH^z  \\
            -(\epsilon_k+H^z)\tilde{H}^x & (\epsilon_k-H^z)\tilde{H}^x &
               2H^x\tilde{H}^x \\
            (\epsilon_k-H^z)\tilde{H}^x & -(\epsilon_k+H^z)\tilde{H}^x &
               2H^x\tilde{H}^x
         \end{array}
         \right).        \nonumber
\end{eqnarray}
Taking the first row
of $\bf L$ from Eq.~\ref{lreqn} and the inhomogeneity vectors for
$mn=\{0,1\}$ from Table~\ref{t:canda}, we find from the regularity condition
 ${\bf L}^0{\bf A}={\bf 0}$,
\begin{equation}
   \frac{\tilde{H}^x}{H^z} = \frac{\la S^x \ra}{\la S^z \ra}.
   \label{eregcon}
\end{equation}
This implies that the ratio on the lhs is not dependent
upon the momentum. This can also be seen directly
from the definitions in Eq.~\ref{diagdef}:
\begin{equation}
  \frac{\la S^x \ra}{\la S^z \ra}=\frac{B^x}{B^z+Dq\la S^z \ra}.
\end{equation}
From the same definitions it is clear that the ratio
$\frac{H^x}{H^z}$ {\em does} depend upon momentum because $H^x$ differs
from $\tilde{H}^x$ by a momentum-dependent term; hence the projector
${\bf R}^0{\bf L}^0$ onto the null space is momentum dependent.

The singular vectors in $\bf U$ and $\bf V$ (see Eq.~\ref{esvdnz})
may be obtained as the
eigenvectors of the symmetric matrices $\mathbf \Gamma \tilde{\mathbf \Gamma}$
and $\tilde{\mathbf \Gamma}\mathbf \Gamma$, respectively. The singular
values in $\bf W$ are the square roots of the eigenvalues of these
matrices.
\begin{eqnarray}
 \bf W &=& \left(
           \begin{array}{ccc}
             W_{11} & 0 & 0 \\
             0 & W_{22} & 0 \\
             0 & 0 & 0
           \end{array}
           \right), \\
  W_{11} &=& \sqrt{H^zH^z+2H^xH^x},  \\
  W_{22} &=& \sqrt{H^zH^z+\frac{1}{2}\tilde{H}^x\tilde{H}^x}, \\[2mm]
 \bf U &=& \left(  %\large
           \begin{array}{cc|c}
             -\frac{1}{\sqrt{2}}
            & \frac{-H^z}{\sqrt{2}W_{22}}
            & \frac{\tilde{H}^x}{2W_{22}} \\[2mm]
%           & \frac{-H^z}{\sqrt{2H^zH^z+\tilde{H}^x\tilde{H}^x}}
%           & \frac{\tilde{H}^x}{\sqrt{4H^zH^z+2\tilde{H}^x\tilde{H}^x}} \\
              \frac{1}{\sqrt{2}}
            & \frac{-H^z}{\sqrt{2}W_{22}}
            & \frac{\tilde{H}^x}{2W_{22}}   \\[2mm]
%           & \frac{-H^z}{\sqrt{2H^zH^z+\tilde{H}^x\tilde{H}^x}}
%           & \frac{\tilde{H}^x}{\sqrt{4H^zH^z+2\tilde{H}^x\tilde{H}^x}} \\
               0
            & \frac{\tilde{H}^x}{\sqrt{2}W_{22}}
            & \frac{H^z}{W_{22}}
%           & \frac{4\tilde{H}^x}{\sqrt{2H^zH^z+\tilde{H}^x\tilde{H}^x}}
%           & \frac{2H^z}{\sqrt{4H^zH^z+2\tilde{H}^x\tilde{H}^x}}
           \end{array}
           \right), \label{umono}\\[4mm]
 \tilde{\bf V} &=& \left( %\large
           \begin{array}{ccc}
              \frac{-H^z}{\sqrt{2}W_{11}}
            & \frac{-H^z}{\sqrt{2}W_{11}}
            & \frac{2H^x}{\sqrt{2}W_{11}}     \\[2mm]
%             \frac{-H^z}{\sqrt{4H^xH^x+2H^zH^z}}
%           & \frac{-H^z}{\sqrt{4H^xH^x+2H^zH^z}}
%           & \frac{H^x} {\sqrt{4H^xH^x+2H^zH^z}} \\
              \frac{-1}{\sqrt{2}} & \frac{1}{\sqrt{2}} & 0 \\[2mm]
              \hline
            & & \\[-4mm]
              \frac{H^x}{W_{11}}
            & \frac{H^x}{W_{11}}
            & \frac{H^z}{W_{11}}
%             \frac{H^x}{\sqrt{2H^xH^x+H^zH^z}}
%           & \frac{H^x}{\sqrt{2H^xH^x+H^zH^z}}
%           & \frac{H^z}{\sqrt{2H^xH^x+H^zH^z}}
           \end{array}
           \right).  \label{evtild}
\end{eqnarray}
The matrix $\bf U$ has been blocked into $\bf u$ and ${\bf u}_0$ by
the vertical line and $\tilde{\bf V}$ into $\tilde{\bf v}$ and
$\tilde{\bf v}_0$ by the horizontal line. We see here explicitly
(by factoring out $H^z$ and applying the regularity condition) that
the vector ${\bf L}_0$, the first row-vector in the matrix ${\bf L}$ in
Eq.~\ref{lreqn}, is proportional to the vector ${\bf u}_0$ and is
independent of momentum. In fact, the vectors in $\bf u$ are also
independent of momentum. Similarly, ${\bf R}_0$ is proportional to
${\bf v}_0$, but {\em does} depend upon momentum. As the
momentum varies, $\tilde{\bf v}_0$ also varies but
remains
in the plane containing the $G^z$-axis and the line bisecting the
axes $G^+$ and $G^-$.  This implies that one of
the vectors in the non-null space can be chosen to be perpendicular
to this plane and therefore must be {\em independent} of momentum. This is
the second row-vector in $\tilde{\bf V}$ in Eq.~\ref{evtild}. The first
row-vector in $\tilde{\bf V}$ is also dependent upon
momentum and
lies in the same plane as the null vector.

From these considerations, it is clear that,
for the monolayer, only the second row-vector of $\tilde{\bf V}$
in Eq.~\ref{evtild}
is {\em momentum-independent}; hence, only
the second row from Eq.~\ref{smallc}
(here specialized to the monolayer) can be used in the consistency
equations.
For spin $S=1$, there are two
expectation values which act as the variables which must be iterated to
self-consistency in Eq.~\ref{eworking}, $\la S^z \ra$ and
$\la S^zS^z\ra$. Since only one row of Eq.~\ref{smallc} can be used, it
is necessary to use this equation twice, each time with different values
of $mn$ (see Table~\ref{t:canda}). The other correlations in
Table~\ref{t:canda} can be expressed in terms of
$\la S^z \ra$ and $\la S^zS^z\ra$ via the regularity condition
${\bf L}^0{\bf A}^{mn}={\bf 0}$, for $m+n \leq 2S+1$
(for details see \cite{FJKE00}).

In the multilayer case, each layer supplies an extra row in
Eq.~\ref{smallc} which can be used as a consistency equation, so that
there are just enough equations to solve for the expectation values
$\la S^z_i \ra$ and $\la S^z_iS^z_i\ra$ for each layer $i$.

\subsubsection{Algebraic properties of the multilayers}
While it is not practicable to perform the singular value decomposition
algebraically for the multilayer films, it is nevertheless possible
to deduce some properties of the singular vectors by examining the
results for the monolayer and combining these with the structure of
the $3\times 3$ blocks of the matrix $\mathbf \Gamma$ in Eq.~\ref{bigamma}.

We start with the structure of the null vectors of the matrix
$\mathbf \Gamma$ from Eq.~\ref{gamthree} in a film of $N$ layers. We
may write a null row-vector in terms of the three components for each layer as
\begin{equation}
   \tilde{\bf v}_0 = (\tilde{v}_0^{(1)},\ldots,\tilde{v}_0^{(N)})
   \label{vnull}
\end{equation}
There are $N$ null vectors each with $3N$ components.
We see from Eq.~\ref{evtild} that the components $\tilde{v}_0^{(i)}$ of
the null vector for the $i$th layer {\em considered in isolation} have
the form $(a_i,a_i,b_i)$. But this must also hold for each of the null vectors
in Eq.~\ref{vnull}.  To see this, consider the $\mathbf\Gamma$-matrix
of Eq.~\ref{gamthree} extended to $N$ layers acting on a single null vector:
\begin{equation}
   {\mathbf\Gamma} {\bf v}_0 = \bf 0
\end{equation}
This gives $3N$ equations for the components of ${\bf v}_0$, 3
for each layer.
Denote the components belonging to layer $i$ as $(x_i,y_i,z_i)$
Now the structure of the $3\times 3$ blocks in $\mathbf\Gamma$ is such
that the {\em first} equations for each layer
(i.e. equations $1,4,7,\ldots$)
involve only the components $x_i$ and $z_i$ for all $i$; i.e. the $N$
equations suffice to determine the $N$ ratios $x_i/z_i$. The {\em second}
equations for each layer (i.e. $2,5,8,\ldots$) involve only $y_i$ and
$z_i$ and they are {\em exactly the same} as the first equations for
each layer. The components $x_i$ are therefore the same as the $y_i$.
But this implies, from the structure of the blocks in $\mathbf\Gamma$,
that the {\em third} equations for each layer are
then automatically satisfied.  The values of the $x_i$, $y_i$, and
$z_i$ are then obtained from normalization.
This means that
each null vector, although perhaps dependent upon the momentum, must
lie in the hyperplane in which the components belonging to layer $i$
have the {\em form} $(a_i,a_i,b_i)$. This, in turn, means that the hyperplane in
which {\em all} layer components have the form $(c_i,-c_i,0)$ must lie in the
non-null space and be independent of momentum,
since each vector of this form
is obviously orthogonal to each null vector.

Consider now special vectors
of this form for which the components for layer $i$ are $(c_i,-c_i,0)$ but
all other components are zero, i.e. $N$ of the $3N$ reference vectors
of Eq.~\ref{erefvec},
$\tilde{\bf V}_{\rm ref}$. Clearly, these vectors are also in the
non-null space and are orthogonal to each other. This is the reason why
the labelling procedure picks out these particular reference vectors as
being identical with some of the vectors $\tilde{\bf v}$. Of course, the
rest of the vectors $\tilde{\bf v}$ are not in general identical to any
of the remaining $2N$ reference vectors, because the null vectors do not
in general have a layer-structure, implying that each of the remaining
reference vectors lies partially in the null space. All of these
features are fully verified by the numerical calculations of the
next section.

In summary, we see that in the $N$-layer film, there are only $N$
components of the vector $\bf c$ from Eq.~\ref{smallc} (for a given
value of $mn$) which are independent of momentum {\em and also} possess
a layer-structure. The numerical labelling procedure is capable of
delivering these vectors. This bit of serendipity
is essential to the success of the entire method, since we
only know how to determine the {\em diagonal} correlations, $C_{ii}$,
from the expectation values; hence, we require that the dot product
of our vector with $C_i({\bf k})$ only involve the component $C_{ii}$.  This
is ensured by the aforementioned property of those (labelled) vectors in
$\tilde{\bf v}$ which are independent of momentum. In contrast,
the momentum-dependent
vectors in $\tilde{\bf v}$ do not share this property and therefore
cannot be used in solving the consistency equations
(\ref{eworking}).

\subsection{Numerical results}
We conclude our exposition with a few illustrative numerical calculations
for the $S=1$ Heisenberg Hamiltonian model
with exchange anisotropy of Eq.~\ref{ehamil}.
The exchange energy is $J=100$, the exchange anisotropy is $D=0.7$, and
the field in the $z$-direction is zero. We show, as representative
examples, some results of calculations
for 1-, 3-, and 7-layer films.

First, the labelled and smoothed row-vectors
$\tilde{\bf V}$ (${\bf V}=({\bf v},{\bf v_0})$) are shown in
Table~\ref{t:vecs} for a 3-layer film with $B^x=0.1,B^z=0$ at $T=107$,
which is very near the reorientation temperature of the
magnetization, $T_R$.
Displayed are the six
vectors in the non-null space, $\tilde{\bf v}$, and the three
null vectors, $\tilde{\bf v}_0$, at $k=0.0100113$.
\begin{table}
   \vspace{0cm}
   \caption{\label{t:vecs}Row-vectors $\tilde{\bf V}$ at
   $k=0.0100113$ for the 3-layer film.}
\begin{tabular}{|c|ccc|ccc|ccc|}\hline\hline
Layer
  &       &       1   &           &           &       2 & & & 3 & \\
$\alpha$
  & + & - & z & + & - & z & + & - & z    \\ \hline
  & 0.71 &    -0.71 &       0   &       0   &       0   &       0   &       0   &       0   &       0    \\
  & 0.46 &     0.46 &    -0.76 &     0.03 &     0.03 &     0.04 &     0.01 &     0.01 &     0.02   \\
$\tilde{\bf v}$  &
      0   &       0   &       0   &     0.71 &    -0.71 &       0   &       0   &       0   &       0    \\
  & 0.02 &     0.02 &     0.02 &     0.47 &     0.47 &    -0.74 &     0.02 &     0.02 &     0.02  \\
  &   0   &       0   &       0   &       0   &       0   &       0   &     0.71 &    -0.71 &       0    \\
  & 0.01 &     0.01 &     0.02 &     0.03 &     0.03 &     0.04 &     0.46 &     0.46 &    -0.76  \\
  \hline
  & 0.54 &     0.54 &     0.65 &    -0.02 &    -0.02 &     0.025 &    -0.01 &    -0.01 &     0.02  \\
$\tilde{\bf v}_0$  &
    0.02 &     0.02 &    -0.04 &    -0.53 &    -0.53 &    -0.67 &     0.02 &     0.02 &    -0.04  \\
  & 0.01 &     0.01 &    -0.02 &     0.02 &     0.02 &    -0.03  &    -0.54 &    -0.54 &    -0.65  \\
  \hline\hline
\end{tabular}
\end{table}
\begin{figure}
\begin{center}
\protect
\includegraphics*[bb=0 0 697 451,
angle=0,clip=true,width=8cm]{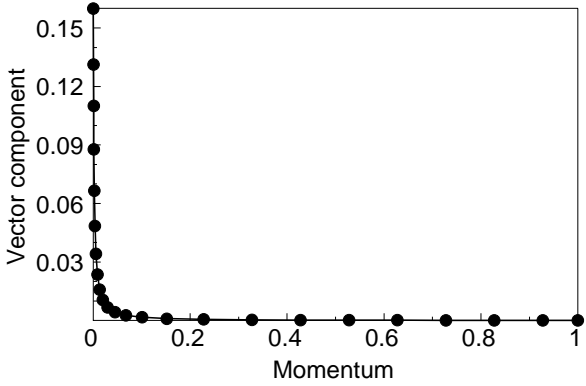}
\protect
\caption{\label{f:refpoint}
One component of a momentum-dependent vector $\tilde{\bf v}$ for the
3-layer film near the Reorientation temperature of the
magnetization ($T=107$)
at the reference points $k_i$.}
\end{center}
\end{figure}
Vectors 1,3, and 5 are independent of momentum and have a layer-structure:
i.e. they may be {\em identified} with layers 1,2, and 3, respectively.
In contrast to this, vectors 2, 4, and 6 have contributions from each
layer. Nevertheless, each of these vectors can be
{\em labelled} by the layers
1,2, and 3, respectively, because the contributions from those layers
are by far the greatest in each respective vector. The null-vectors
are similar to vectors 2, 4, and 6: they
vary with $k$, have no layer-structure, but may be labelled with a
layer index.

A component of one of the vectors $\tilde{\bf v}$ is shown in
Fig.~\ref{f:refpoint} at the
reference points $k_i$ for one of the momentum-dependent
vectors for the 3-layer film at a temperature $T=107$
near the reorientation temperature of the magnetization.
Since there is rapid variation near $k=0$, the number of reference
points there is denser than at larger  $k$. At smaller temperatures,
this behaviour is not as extreme.

\begin{figure}
\begin{center}
\protect
\includegraphics*[bb=0 0 697 451,
angle=0,clip=true,width=11cm]{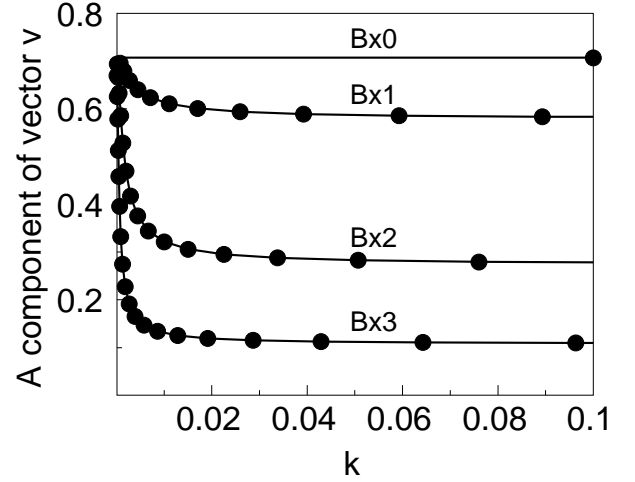}
\protect
\caption{\label{f:vcomp}
A single component of one of the vectors in $\tilde{\bf v}$ for the monolayer
as a function
of momentum $k$ for increasing values of magnetic field $B^x$:
$B^{x_0}\approx 0 < B^{x_1}\ldots < B^{x_3} \approx 1.3$. The circles
indicate the position of the reference vectors.}
\end{center}
\end{figure}

Fig.~\ref{f:vcomp} displays the most rapidly varying component of one of
the momentum-dependent labelled vectors $\tilde{\bf v}$ in the non-null space
as a function of the momentum $k$ over
the first tenth of the $k$-range for the monolayer at $T=0$;
the curves remain roughly constant over the
rest of the range. For increasing magnetic field in the $x$-direction,
$B^{x_i}$, the curves become steeper near $k=0$, becoming nearly L-shaped
for the largest field value (corresponding to $\la S^z \ra \rightarrow 0$).
A greater density of reference vectors (positions are indicated by the
full circles) is needed in this range.
Nevertheless, the smoothing procedure is able to deliver acceptable
vectors even under these extreme conditions. Note that $all$ the vectors
$\tilde{\bf v}$ in the non-null space are needed accurately, not just
those that are independent of momentum, because they are employed in reducing
the size of the $\mathbf \Gamma$-matrix by the transformation in
Eqns.~\ref{gamtrans} to~\ref{ctrans}.

\begin{figure}
\begin{center}
\protect
\includegraphics*[bb=85 75 538 355,          %0 0 697 451,
angle=0,clip=true,width=9cm]{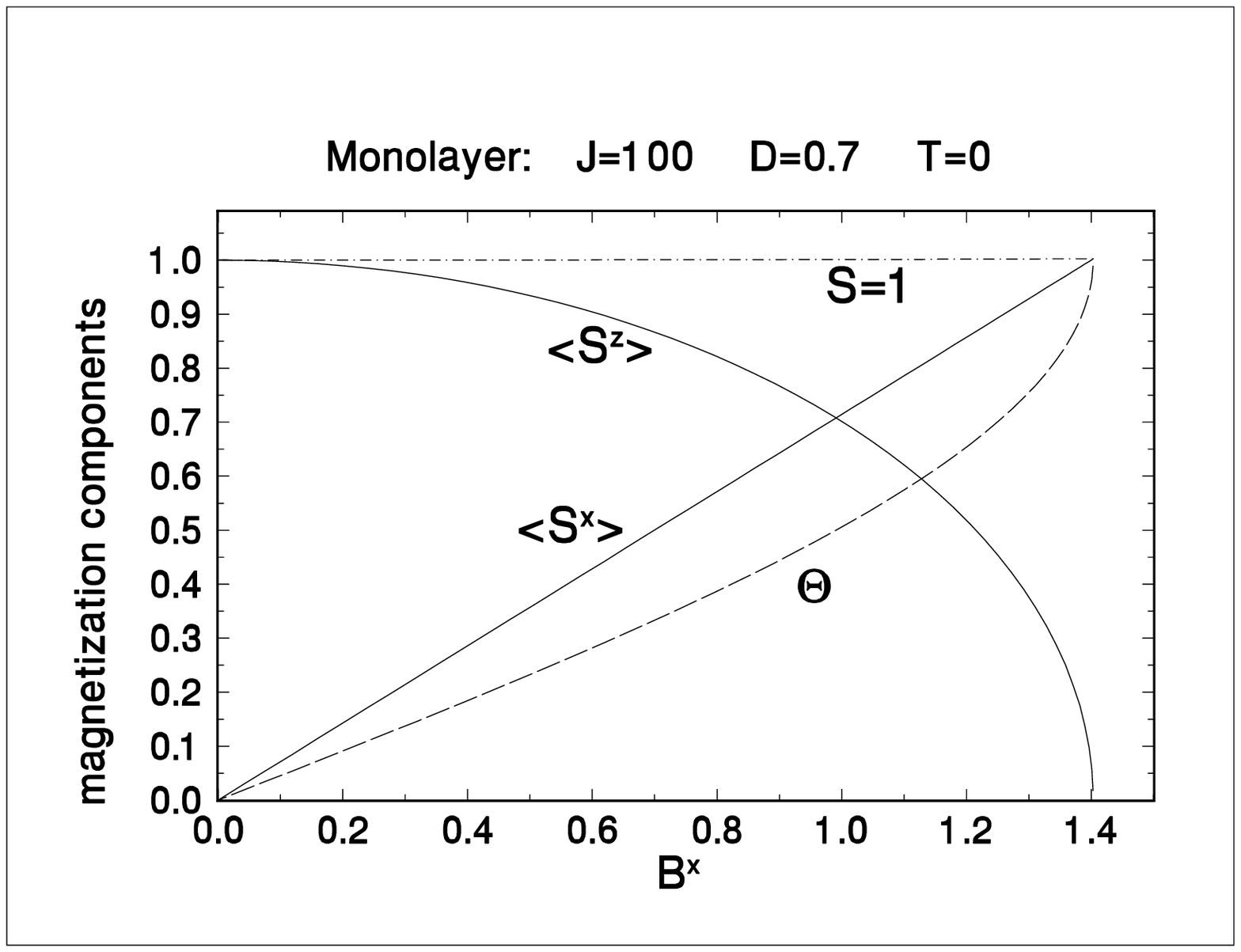}   %{gb01-001.eps}
\protect
\caption{\label{f:magvonbx}
Magnetization components $\la S^z\ra$ ($=1$ at $B^x=0$)
and $\la S^x\ra$ (solid lines),
the reorientation angle of the magnetization,
$\Theta=(\arctan{\frac{\la S^x \ra}{\la S^z \ra}})/\frac{\pi}{2}$
(dashed line),
and the absolute value of the total magnetization (dash-dot chain)
for a spin S=1 Heisenberg monolayer as functions of the
applied magnetic field in the $x$-direction, $B^x$, at $T=0$.}
\end{center}
\end{figure}

We now present the magnetizations calculated by solving Eq.~\ref{eworking}
self-consistently as a function of magnetic field and temperature.
Fig.~\ref{f:magvonbx} shows the
magnetization $\la S^z\ra$ (beginning at 1 at $B^x=0$), the magnetization
$\la S^x\ra$ increasing linearly from 0, the reorientation
angle of the magnetization, $\Theta$,
and the absolute value of the total magnetization
for a spin $S=1$ Heisenberg monolayer as functions of the
applied magnetic field in the $x$-direction, $B^x$ at temperature $T=0$.
(Results at other temperatures are similar.)

\begin{figure}
\begin{center}
\protect
\includegraphics*[bb=85 75 538 355,          %0 0 697 451,
angle=0,clip=true,width=9cm]{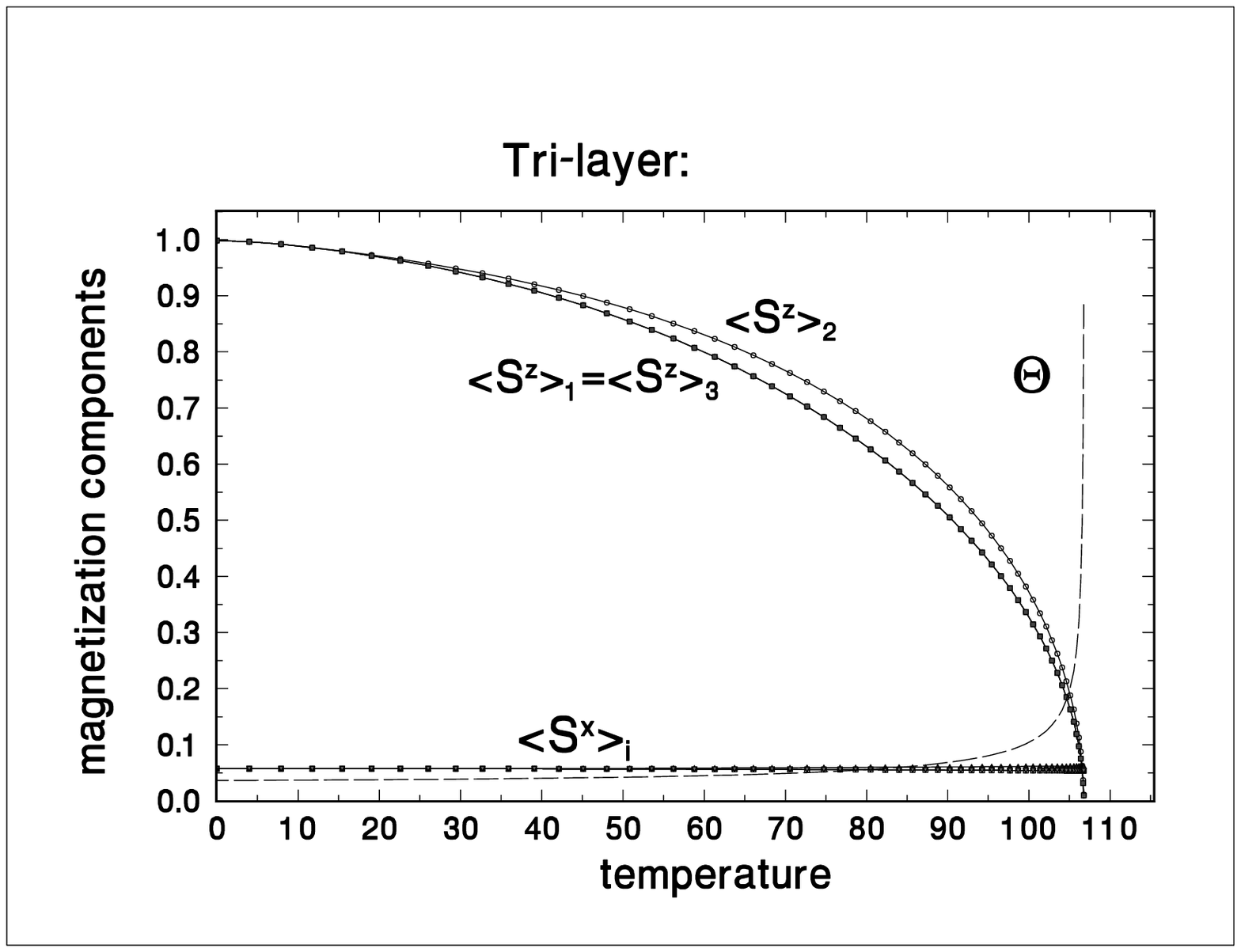}
\protect
\caption{\label{f:magvont}
Magnetization components in different layers
$\la S^z\ra_i$ and $\la S^x\ra_i$
(solid lines) and the reorientation angle of the magnetization
(dashed line),
for a spin $S=1$ Heisenberg trilayer as functions of the
temperature, $T$, at $B^x=0.1$}
\end{center}
\end{figure}

Magnetization components as a function of temperature
are shown for the 3-layer film
in Fig.~\ref{f:magvont}. Magnetizations for layers~1 and~3 are
identical as required by symmetry. Because the magnetic field
component ${B}^x$ is so small, the 3 components in the $x$-direction
cannot be distinguished from one another on the scale of the diagram.

\begin{figure}
\begin{center}
\protect
\includegraphics*[bb=150 87 544 691,
angle=-90,clip=true,width=9cm]{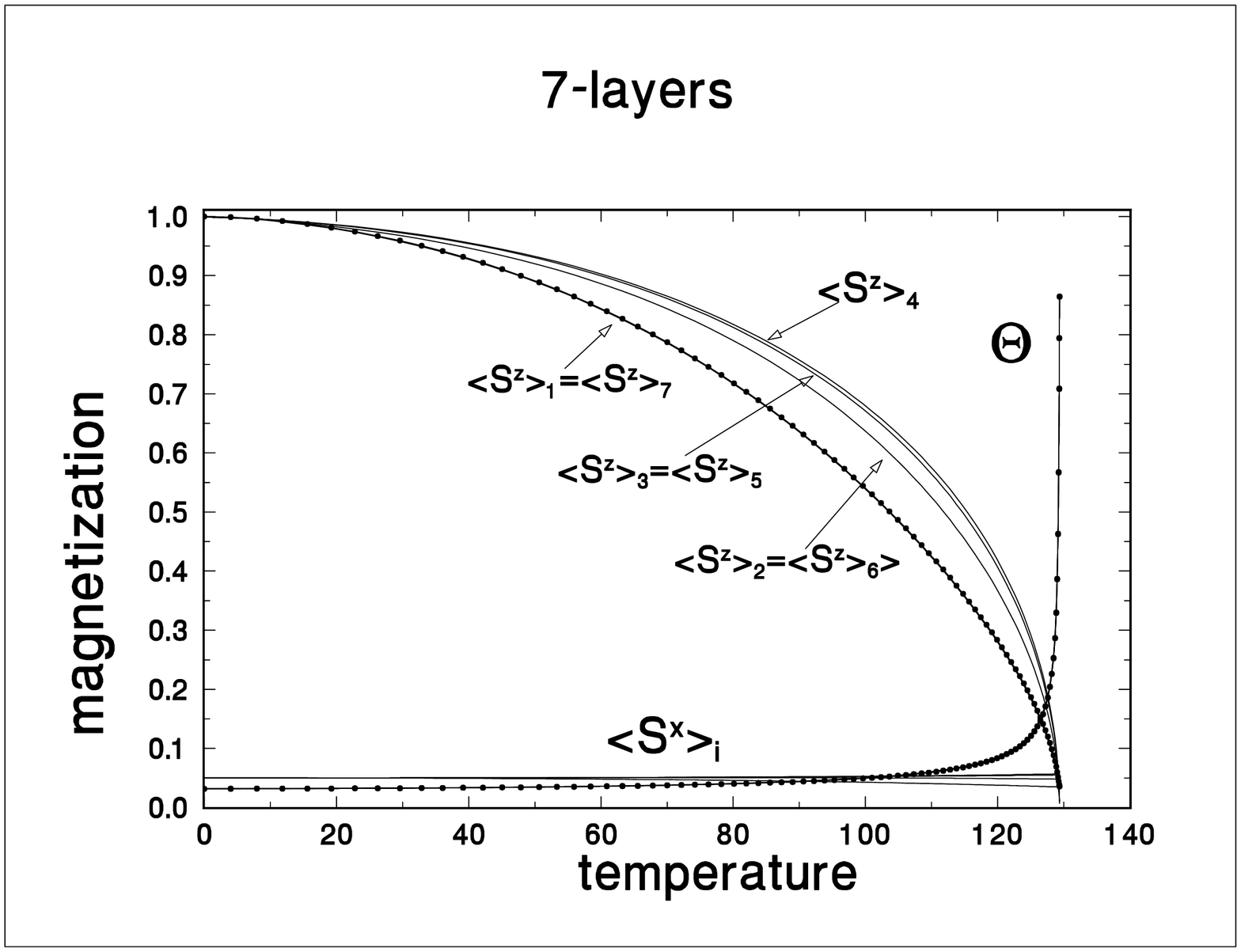}
\protect
\caption{\label{f:seven}
Magnetization components in different layers
$\la S^z\ra_i$ and $\la S^x\ra_i$
(solid lines) and the reorientation angle of the magnetization
(dashed line),
for a spin $S=1$ Heisenberg $7$-layer film as functions of the
temperature, $T$, at $B^x=0.1$}
\end{center}
\end{figure}

Fig.~\ref{f:seven} demonstrates that it is also no problem to calculate
the reorientation of the magnetization for a film consisting of
$7$~layers. Films with spin $S>1$ can also be treated. In fact, all of
our previous work on this model\cite{FK02}
can be reproduced with the present method, so there is no need to
present more examples here.

\section{A more efficient algorithm}
In this section, we shall establish the connection between the
method developed here and the one we used in our previous
work on ferromagnetic Heisenberg films with exchange
anisotropy\cite{FK02}.
To do this, the working equations are cast in a slightly
different form that in some cases allows one to dispense with
the smoothing procedure. This suggests that the new method serves
not only as a tool to find appropriate consistency
equations but also as a means of designing an efficient algorithm
for solution.

Our method has so far employed the matrix of
singular vectors $\tilde{\bf v}$
to transform the correlations in momentum space
(${\bf c}=\tilde{\bf v}{\bf C_{k}}$) so that  ${\bf c}$ can be
expressed via Eq.~\ref{esvdcor} in terms of quantities referring
only to the non-null space of the matrix $\mathbf \Gamma$.
Via the labelling procedure, the row-vectors $\tilde{\bf v}$
were then transformed so that a subset of them were independent of
momentum; hence, the result in Eq.~\ref{ecterm} could be exploited
in the consistency equations Eq.~\ref{eworking}. Let us now
multiply the rhs of Eq.~\ref{esvdcor} by the unit matrix
$\tilde{\bf v}{\bf v}$:
\begin{equation}
   \tilde{\bf v}{\bf C}_{\bf k} = \tilde{\bf v}({\bf v}{\bf r}
         {\cal E}^1{\bf l \tilde{\bf v}A}).
\end{equation}
Consider now the consistency equation obtained from the
momentum-independent vector $\tilde{\bf v}_j$, which we
take as the second row-vector $\tilde{\bf V}$ in
Eq.~\ref{evtild} (refer also to Eq.~\ref{ecterm}):
\begin{equation}
   \tilde{\bf v}_j\int d{\bf k}\ {\bf C}_{\bf k}
   = \tilde{\bf v}_j {\bf C}
   = \int d{\bf k}\ \tilde{\bf v}_j({\bf v}{\bf r}
         {\cal E}^1{\bf l \tilde{\bf v}A}).
         \label{ehybrid}
\end{equation}
There is one such equation for each layer. Taken together, these
equations succinctly describe the method of solution in
reference \cite{FK02}, where, for each layer,
a difference of correlations (the lhs) was related to the
difference between the first and second integrated components of the
rhs of the equation in round brackets. The idea to build this
difference of components came from the empirical observation
(by taking linear combinations of the equations of the standard
procedure in the full space)
that such difference terms make no contribution to
the momentum-dependent term
${\bf R}^0{\bf L}^0$, thus eliminating the latter from the
consistency equations. Here (Eq.~\ref{ehybrid}), taking the
difference is effected via multiplication by the vectors
$\tilde{\bf v}_j$ found systematically from the SVD coupled
with the labelling procedure, thus replacing
intuition with a systematic approach.

The advantage of this way of solving the
equations is that in this case there
is no need to apply the smoothing procedure at each value of
$\bf k$, since the singular vectors $\bf v$ and $\tilde{\bf v}$ now
{\em both} occur in the product in the integrand; i.e. an arbitrary
sign arising from $\tilde{\bf v}$ is offset by that from $\bf v$.
We emphasize that smoothing would still be needed if there were
degeneracies in the non-zero singular values, which was not the case
in our previous work. Also, the labelling and smoothing procedures
are {\em required} to {\em find} the appropriate
$\tilde{\bf v}_j$ in the diagnostic phase of the work.

This suggests that the procedure in this paper can be used as a
tool to find appropriate vectors $\tilde{\bf v}_{LS}$ via
the labelling and smoothing procedures and then, {\em once the
vectors have been found}, to design a more
efficient numerical procedure.
We note here that the momentum-independent vectors for
the exchange anisotropy model are particularly simple, being in fact
just numbers.  This need not always be the case: in general, these vectors
will depend on the magnetizations and will vary as one moves
through solution space (e.g. magnetizations as a function of
temperature). In this case, a method of the type suggested
above would require that the diagnostic tool be built into the
numerical algorithm in order to find the appropriate vectors
at each point in space.

\section{Discussion}
The main thrust of this paper is to demonstrate that the SVD is a
very powerful tool for solving the problems which arise when the
equations of motion matrix $\mathbf \Gamma$ has zero eigenvalues.
In particular, the singular vectors in the non-null space, $\tilde{\bf v}$,
are used to {\em reduce the size} of the matrix that needs to be
diagonalized and, at the same time, {\em increase the numerical
stability} by eliminating the zero eigenvalues and the
degeneracy associated with them, a fact of considerable practical
importance, especially when some of the non-zero eigenvalues lie
{\em close} to zero. In addition, the suitably labelled and smoothed
singular vectors $\tilde{\bf v}_{LS}$
 serve to transform the consistency
equations so as to deliver the correlation functions in coordinate
space required for the solution of these equations.

The method outlined
here can be viewed as a recipe for direct numerical calculation .
The recipe is quite simple:
\begin{enumerate}
   \item At the current point in solution space, generate
a sufficient number of reference vectors $\tilde{\bf V}_{\rm ref}(k_i) $ at
suitably chosen knots $k_i$. \\
   \item  At each desired value of $k$,
\begin{enumerate}
   \item Obtain the untreated singular vectors $\tilde{\bf v}$ in the
         non-null space from the SVD
         of the $\mathbf \Gamma$-matrix.  \\
   \item If necessary, find labelled vectors from these:
         $\tilde{\bf v}_L=\tilde{\bf Z}\tilde{\bf v}$.
         If no labelling is desired, set $\tilde{\bf Z}=1$ at this point. \\
   \item Find the set of smoothed reference vectors appropriate to $k$
         and carry out the smoothing transformation:
         $\tilde{\bf v}_{LS}
               ={\cal L}\tilde{\cal Z}\tilde{\bf Z}\tilde{\bf v}$.
         \\
   \item Use the smoothed vectors to reduce the $\mathbf \Gamma$-matrix by
         eliminating the null space. Diagonalize the reduced matrix to get
         the eigenvalues and eigenvectors:
         $\bf l{\mathbf \gamma}\bf r={\mathbf \omega}^1$ \\
   \item Select the appropriate (i.e. properly labelled
         momentum-independent)
         vectors $\tilde{\bf v}$ to generate the set of consistency
         integrands of Eq.~\ref{eworking}:
         ${\bf r}{\cal E}{\bf l}\tilde{\bf v}{\bf A}
         -\tilde{\bf v}{\bf C}_{\bf k}$  \\
\end{enumerate}
   \item Finally, solve the consistency equations
Eq.~\ref{eworking} (e.g. with the method described in \cite{FKS02}).
\end{enumerate}

 This recipe cannot be so general as to apply to {\em any}
problem, for the labelling is dictated by the properties of the
$\mathbf \Gamma$-matrix, which will depend upon the nature of the
physical model and the approximations used in the decoupling leading to
the $\mathbf \Gamma$-matrix. The method can also fail if it is not
possible to find momentum-independent vectors.

The method may also be useful as a diagnostic tool to aid
in the design of an efficient computation method. Once the consistency
equations have been found by our systematic
procedure, Eq.~\ref{ehybrid} can be used as a more efficient
working equation, since it usually does not require the smoothing
procedure. In other words, our procedure, used
diagnostically, serves to identify the {\em allowed} consistency
equations. We are then free to manipulate these to form a new set
of allowed equations having simpler numerical requirements,
thus leading to an
optimal solution of any particular solvable problem.

\appendix
\section{Reduction of the $\mathbf\Gamma$-matrix
         with the Singular Value Decomposition (SVD)}
In Eq.~\ref{gamtrans}, the matrix $\mathbf\Gamma$ is reduced in
dimension by the number of zero eigenvalues, $N_0$, via a transformation
by the singular vectors spanning the non-null space, $\tilde{\bf v}$.
It is very important that the eigenvalues of the reduced matrix,
$\mathbf\gamma$, be identical with the non-zero eigenvalues of
$\mathbf\Gamma$, $\omega^1$. In this appendix, we show under which
conditions we can expect this to be the case.

While the null eigenvectors of $\mathbf\Gamma$,
${\bf L}^0$ and ${\bf R}^0$, lie completely in the spaces spanned by
${\bf u}_0\tilde{\bf u}_0$ and ${\bf v}_0\tilde{\bf v}_0$, respectively,
the vectors belonging to the non-zero eigenvalues, ${\bf L}^1$ and
${\bf R}^1$, may have components in both the null space and
non-null space. (e.g. $\tilde{\bf v}{\bf R}^0=0$ but
$\tilde{\bf v}_0{\bf R}^1 \neq 0$ in general). In what follows, we
assume that the left and right eigenvectors of $\mathbf\Gamma$ have
been constructed to be orthonormal: $\bf LR = 1$.

Because $\bf L$ and $\bf R$ diagonalize $\mathbf\Gamma$, we have,
inserting ${\mathbf\Gamma}={\mathbf\Gamma}{\bf v}\tilde{\bf v}$ and
${\bf v}\tilde{\bf v}+{\bf v}_0\tilde{\bf v}_0=\bf 1$,

\begin{eqnarray}
   {\mathbf \Omega} &=& {\bf L}{\mathbf\Gamma}{\bf R}, \nonumber \\
   \left(
   \begin{array}{cc}
    \bf  0 & \bf 0 \\
    \bf  0 & {\mathbf \omega}^1 \\
   \end{array}
   \right)
   & = &
   \left(
   \begin{array}{c}
     \bf L^0 \\
     \bf L^1 \\
   \end{array}
   \right) ({\bf v \tilde{\bf v}}+{\bf v}_0\tilde{\bf v}_0)
           {\mathbf\Gamma}{\bf v \tilde{\bf v}}
           ({\bf R}^0,{\bf R}^1),
   \nonumber \\
   &=&
   \left(
   \begin{array}{cc}
     \bf e_{11} & \bf e_{12} \\
     \bf e_{21} & \bf e_{22} \\
   \end{array}
   \right),
\end{eqnarray}
where the individual matrix blocks are
\begin{eqnarray}
   {\bf e}_{11} &=&
   {\bf L}^0({\bf v \tilde{\bf v}}+{\bf v}_0\tilde{\bf v}_0)
   {\mathbf\Gamma}{\bf v \tilde{\bf v}R}^0,  \nonumber \\
                &=& {\bf L}^0 {\mathbf\Gamma}{\bf R}^0, \nonumber \\
   {\bf e}_{12} &=& {\bf L}^0 {\mathbf\Gamma}{\bf v \tilde{\bf v}R}^1,
                    \nonumber \\
                &=& {\bf L}^0 {\mathbf\Gamma}{\bf R}^1, \nonumber \\
   {\bf e}_{21} &=&
   {\bf L}^1({\bf v \tilde{\bf v}}+{\bf v}_0\tilde{\bf v}_0)
   {\mathbf\Gamma}{\bf v \tilde{\bf v}R}^0,  \nonumber \\
                &=& {\bf L}^1 {\mathbf\Gamma}{\bf R}^0, \nonumber \\
   {\bf e}_{22} &=&
   {\bf L}^1({\bf v \tilde{\bf v}}+{\bf v}_0\tilde{\bf v}_0)
   {\mathbf\Gamma}{\bf v \tilde{\bf v}R}^1.
\end{eqnarray}
Since ${\mathbf\Gamma}{\bf R}^0=0$ and ${\bf L}^0{\mathbf\Gamma=0}$,
the blocks ${\bf e}_{11}$, ${\bf e}_{12}$, and ${\bf e}_{21}$ are zero
as required. The other block is the sum of two terms:
\begin{eqnarray}
   {\bf e}_{22 }&=& {\bf L}^1 {\bf v \tilde{\bf v}}{\mathbf\Gamma}
                    {\bf v \tilde{\bf v}R}^1 +
                    {\bf L}^1 {\bf v}_0 \tilde{\bf v}_0
              {\mathbf\Gamma}{\bf R}^1,  \nonumber \\
      &=& {\bf l}{\mathbf\gamma}{\bf r} +
                    {\bf L}^1 {\bf v}_0 \tilde{\bf v}_0
              {\mathbf\Gamma}{\bf R}^1, \label{ereduced}
\end{eqnarray}
where we have defined ${\bf l}:={\bf L}^1 {\bf v}$ and
${\bf r} = \tilde{\bf v}{\bf R}^1$. If the second term in
Eq.~\ref{ereduced} were zero, then the vectors $\bf l$ and $\bf r$
would diagonalize $\mathbf\gamma$ yielding the
eigenvalues ${\mathbf\omega}^1$
as desired. This condition is fulfilled if the eigenvectors
${\bf R}^0$ span the null space, for we can then express the
vectors ${\bf v}_0$ in terms of the ${\bf R}^0$, each of which is
orthogonal to ${\bf L}^1$ by construction
(${\bf L}^1{\bf R}^0=\bf 0$).  In all of our
calculations, this was indeed the case and was checked numerically
at every reduction of $\mathbf \Gamma$.

\section{Matching of two vector spaces}
In the
context of this paper, we are usually concerned with {\em rotating}
a set of target vectors (which we do not want to disturb in
any other manner)
so as to match as closely as possible a set of (approximate)
reference vectors which
serve as labels or calibration vectors. The method outlined
below treats this problem in a general way and shows how
the SVD can effect this matching in an optimal way. Note that
the notation in this appendix is independent from that in the
main body of the paper.

It often happens that there are two sets of $N$-dimensional vectors
which are related by a general rotation (i.e. a rotation about an
arbitrary axis) and that we wish to find this rotation. If each of the two
sets are subsets of the whole space, it may be that each set
spans a slightly different subspace. This is the problem that we most often
meet here: the two subspaces do not overlap exactly but have a
high degree of overlap. In this case, we wish to find the ``best''
general rotation in the sense that each vector of one set is brought
into as close a coincidence as possible with some vector of the other
set.

Let $r =1,\dots,N_R$ label the $N_R$ column vectors of length $N$,
the {\em reference} vectors,
in the matrix $\bf R$ of dimension $N\times N_R$. Similarly,
let $t =1,\dots,N_T$ label the $N_T$ column vectors of length $N$,
the {\em target} vectors,
in the matrix $\bf T$ of dimension $N\times N_T$. We assume here that
the vectors in $\bf R$ and $\bf T$ are orthonormal among themselves
($\tilde{\bf R}{\bf R}=\bf 1$ and $\tilde{\bf T}{\bf T}=\bf 1$.)
We wish to find
a $rotation$ which, when applied to the
{\em target} vectors, brings them into closest coincidence with the
reference vectors.
In order to ensure that there are no systematic degeneracies in the
singular values that we will calculate,
we first associate the vector $R_{\bullet,i}$ (i.e. the column $i$
of the matrix $\bf R$) with
the unique label $L(i)=N_R+1-i$ (i.e. a label which {\em decreases} as $i$
increases. Then, we define a weighted overlap matrix whose
$N_R\times N_T$ elements are given by
\begin{equation}
  {\bf S}_{rt} := \sqrt{L(r)}\tilde{\bf R}_{r\bullet}{\bf T}_{\bullet t}.
  \label{aisdef}
\end{equation}
The SVD of the matrix $\bf S$ is then
\begin{equation}
   {\bf S}_{N_R \times N_T}={\bf L}_{N_R\times N_T}
   {\bf y}_{N_T\times N_T} \tilde{\bf Z}_{N_T \times N_T}.
   \label{aisvd}
\end{equation}
The matrix of the label operator in the basis of target vectors is
\begin{equation}
   {\bf P}_{N_T \times N_T} = \tilde{\bf S}{\bf S}.
   \label{labelop}
\end{equation}
Writing $\bf S$ in terms of the SVD in this equation shows that the
matrix $\bf Z$ diagonalizes $\bf P$ to produce eigenvalues ${\bf y}^2$,
so that the required rotation is just $\bf Z$. i.e.
\begin{equation}
{\bf T}'= \bf TZ.
\end{equation}
The justification for this transformation is that
the matrix of the label operator in this new basis is diagonal:
\begin{equation}
   {\bf P}'=\tilde{\bf S}'{\bf S}' = {\bf y}^2.
\end{equation}

If $N_R=N_T$ and the reference and target vectors lie in the same
subspace, then the singular values $y_i$ will be identical to
$\sqrt{L(i)}$ provided that the singular values are numbered in decreasing
order, the usual convention; this is the reason for the particular
choice of labels $L(i)$ above. If the reference and target vectors
span slightly different spaces, then there will be some discrepancy
between the singular values and the square roots of the labels.

If the number of reference vectors is less than the number of target
vectors, then some of the singular values of $\bf S$ will be zero. This means
that a subset of the rotated target vectors will be
as closely coincident
with the (smaller) number of reference vectors as possible and the rest of
the (rotated) target vectors will be arbitrary
and span the (null) space of $\bf S$ left over. The
arbitrariness of the vectors is due to the degeneracy of the null space.

If there are more reference vectors than target vectors, then each
(rotated) target vector will be well-defined but the set of target
vectors will not span the space of the reference vectors.

% \newpage

\end{document}